\documentclass[twocolumn,showpacs,preprintnumbers,superscriptaddress,amsmath,amssymb]{revtex4}

\usepackage{graphicx}
\usepackage{dcolumn}
\usepackage{bm}

\newcommand{\SSS}{\hat{S}}
\newcommand{\lefts}{{[}}
\newcommand{\rights}{{]}}
\newcommand{\thalf}{\tfrac{1}{2}}
\renewcommand{\bbox}[1]{\bm{#1}}

\begin{document}


\title{Effective pseudopotential for energy density functionals with higher order derivatives}

\author{F. Raimondi}
 \email{francesco.raimondi@jyu.fi}
\affiliation{%
Department of Physics, P.O. Box 35 (YFL)
    FI-40014 University of Jyv\"askyl\"a, Finland.
}%

\author{B. G. Carlsson}%
\affiliation{%
Department of Physics, P.O. Box 35 (YFL)
    FI-40014 University of Jyv\"askyl\"a, Finland.
}%
\affiliation{Department of Physics, Lund University,
 P.O. Box 118
Lund
22100,
Sweden.}

\author{J. Dobaczewski}
\affiliation{%
Department of Physics, P.O. Box 35 (YFL)
    FI-40014 University of Jyv\"askyl\"a, Finland.
}%
\affiliation{Institute of Theoretical Physics, Faculty of Physics, University of Warsaw,
ul. Ho\.{z}a 69, PL-00-681 Warsaw, Poland.}

\date{\today}

\begin{abstract}
We derive a zero-range pseudopotential that includes all possible
terms up to sixth order in derivatives. Within the Hartree-Fock
approximation, it gives the average energy that corresponds to a
quasi-local nuclear Energy Density Functional (EDF) built of
derivatives of the one-body density matrix up to sixth order. The
direct reference of the EDF to the pseudopotential acts as a
constraint that divides the number of independent coupling
constants of the EDF by two. This allows, e.g., for expressing the isovector
part of the functional in terms of the isoscalar part, or {\it vice
versa}. We also derive the analogous set of constraints for the
coupling constants of the EDF that is restricted by spherical,
space-inversion, and time-reversal symmetries.
\end{abstract}

\pacs{21.60.Jz, 21.30.Fe, 71.15.Mb}

\maketitle

\section{\label{sec:Intro}introduction}

One of the big challenges of the current research in nuclear
structure physics is the search for a universal energy density
functional (EDF) \cite{SCIDAC}. Among different possible approaches
to this search, the consideration of a local or quasi-local EDF based
on the density-matrix expansion (DME) is in recent years the object
of intense studies
\cite{Carlsson1,CarlssonPRL,Dobaczewski1,Gebremariam1,Stoitsov1,Erler1}.
These aim at improving the classic work of
Negele and Vautherin \cite{Negele1,Negele2} and better theoretical
understanding based on the effective theory \cite{[Lep97],[Bog10]}
and on the framework of the density functional theory
\cite{[Sto91]}.

In the recent work~\cite{Carlsson1}, we proposed a new expansion of
the nuclear energy density in higher-order derivatives of densities.
There, following the effective-theory approach, a Skyrme-like
quasi-local next-to-next-to-next-to-leading order (N$^{3}$LO) EDF was
derived with terms of the EDF constrained only by symmetry
principles. In the present derivation, we took the route in opposite
direction as compared to what has been done for the standard Skyrme
next-to-leading order (NLO) EDF. Namely, historically, the Skyrme
force has been initially proposed first as an expansion of the
effective interaction in relative momenta up to second
order~\cite{Skyrme1,Skyrme2}. Next, for this force the average
Hartree-Fock (HF) energy was evaluated, giving the Skyrme EDF with
half of the coupling constants constraint to the other half, see
Ref.~\cite{Perlinska1} for the modern complete analysis. Only later,
a possibility of releasing these constraints was considered and
studied, see, e.g., Ref.~\cite{[Rei95]} for the analysis of the
spin-orbit term.

In the present work we complete the results of Ref.~\cite{Carlsson1}
by deriving the expansion of the effective interaction in relative
momenta up to N$^{3}$LO. This generalizes the Skyrme force up to
sixth order and allows us to make a link with the general N$^{3}$LO
EDF derived in~\cite{Carlsson1}. One should stress that the present
analysis is not at all an independent repetitive derivation of the
same functional. Indeed, the constraints on the EDF coupling
constants, which are induced by the HF averaging of this generalized
force, cannot be obtained without following the path presented in
this study.

The complete higher-order EDFs or pseudopotentials have never yet
been applied in practical calculations. The work towards this goal is
now in progress, with basic derivations like the ones of
Ref.~\cite{Carlsson1} and in the present work coming first, the
construction of numerical codes like the one in Ref.~\cite{[Car10d]}
coming next, and the full adjustments of coupling constants that
will follow. In this respect, at present, we are in a similar phase
of studies as before the chiral N$^{3}$LO potentials for two-nucleon
systems were adjusted, see, e.g., Ref.~\cite{[Epe05]}, and after the
tools for calculating the corresponding N$^{3}$LO diagrams were
developed, see, e.g., Ref.~\cite{[Kai01]}. Nevertheless, studies of
particular higher-order EDF terms have already been
performed~\cite{[Zal10a],[Fan11]}.

In Ref.~\cite{[Car10e]}, the question of convergence of the series in
higher-order derivatives was recently addressed within the DME
applied to the Gogny non-local functional, and it was shown that
every next order up to sixth gives contributions smaller by large
factors. This gives us confidence that fits of higher-order EDFs have
a fair chance of converging. A rigorous power counting scheme,
analogous to what has been introduced in the chiral perturbation
theory \cite{[Wei90b]}, would have to use derivatives of regularized
zero-range interactions, see, e.g.~Ref.~\cite{[Lep97]}. Such a
regularization would provide a proper cut-off scale, against which
the powers of derivatives could be estimated. A good model of the
regularized delta force is the Gaussian interaction, which, however,
leads (through the exchange term) to non-local functionals. Within
the EDF methodology, an effective theory based on derivatives of
finite-range force is in principle possible, but has not yet been
tried because of the degree of numerical complications involved. In
the language of the effective field theory, the power counting scheme
allows us to properly classify diagrams of the perturbation series,
however, the ideas of an effective theory are much more general than
their applications in the field theory -- here we use them within the
framework of standard quantum mechanics of many-body systems.

The EDF description of nuclear states is phenomenological in the
sense that it depends on the coupling constants,
which are usually fitted
to available experimental data, see recent
Refs.~\cite{Klupfel1,Kortelainen1} on fitting the second-order (NLO)
Skyrme functionals. Fits of the full N$^{3}$LO EDF are much more
complicated because of strong inter-dependencies of the coupling
constants and instabilities~\cite{[Kor10a]} occurring in certain
regions of the parameter space. Our main motivation to carry out the
present work was to find constraints on parameters of the general
EDF, which result from its relation to a pseudopotential. Such a
relation reduces the number of parameters that have to be
fit to data, and by this virtue is a positive change, at least at the
preliminary stage of adjustments.

Instead of fitting the coupling constants of the EDF, it is also possible to derive them
directly using the DME~\cite{[Car10e]}. The DME
gives an EDF which approximates more complicated and time consuming
HF calculations based on finite-range forces.
When applying the DME, the relations to pseudopotentials are however
usually broken~\cite{Dobaczewski1}. By enforcing
these relations, as done here, one
ensures that the generated EDF is free from unphysical self-interaction
\cite{[Per81],[Cha10]} and can be applied in beyond-mean-field
applications without problems, see, e.g., Refs.~\cite{[Dug10],[Sat11]}.

By following the standard convention, here we call the generalized
Skyrme force pseudopotential, which is the name denoting a
quasi-local operator depending on spatial derivatives. We also
consequently use the names 'parameters' to denote numerical
coefficients of different terms of the pseudopotential, and we use
the names 'coupling constants' to denote numerical coefficients of
terms in the EDF.

The paper is organized as follows. In Sec.~\ref{sec:the
Pseudopotential} we construct the pseudopotential in two alternative
forms and list all its terms up to N$^{3}$LO. We also evaluate the
constraints imposed by the gauge symmetry. In Sec.~\ref{sec:isoEDF} we
discuss the procedure of HF averaging to obtain the EDF from the
pseudopotential. In particular, in Sec.~\ref{sec:resultsaveraging} we
derive the general relations connecting the parameters of the
Galilean-invariant pseudopotential to the coupling constants of the
EDF, whereas in Sec.~\ref{iso} we derive the constraints for the case
of conserved gauge symmetry. In Sec.~\ref{spherical} we reduce our
results to the case of the conserved spherical, space-inversion, and
time-reversal symmetries. After formulating the conclusions of the
present study in Sec.~\ref{conclusions}, in Appendices
\ref{appA}--\ref{appD} we present derivations related to the
time-reversal invariance and hermiticity of the pseudopotential, we
list results pertaining to the gauge-invariant pseudopotentials, and
we give relations between the two alternative forms of
pseudopotentials. Results obtained in the present work that are too
voluminous to be published in the printed form are collected in the
supplemental material~\cite{suppl}.

\section{\label{sec:the Pseudopotential}General form of the
pseudopotential in the sphe\-rical-tensor formalism}

\subsection{\label{subsec:central-like form of the pseudopotential}
Central-like form of the pseudopotential}

The Skyrme interaction is one of the most important phenomenological
effective interaction used in microscopic nuclear structure
calculations: such two-body interaction is a short-range expansion up
to the second order in derivatives, which contains a certain number
of fit parameters adjusted to reproduce the experimental data. In the
literature the Skyrme interaction is usually written in cartesian
representation, but for our extended pseudopotential we adopt the
spherical-tensor representation of operators \cite{Varshalovich},
whose building blocks can be found in \cite{Carlsson1}.

Depending on the specific form of the coupling of the derivative
operators with the spin operators, different ways to construct the
pseudopotential are possible. A particular form of the
pseudopotential, which we call central-like or LS-like, is
constructed in the present Section. It is based on coupling together
the derivative operators and spin operators, which are then coupled
to rotational scalars. An alternative form, called tensor-like or
JJ-like, is presented in Section \ref{subsec:tensor-like
pseudopotential}. There, each derivative operator is coupled with one
spin operator, and then they are coupled together to rotational
scalars.

In the central-like form, the pseudopotential is a sum of terms,
\begin{equation}
 \hat{V}=\sum_{\substack{\tilde{n}' \tilde{L}', \\ \tilde{n} \tilde{L},v_{12} S}}
      C_{\tilde{n} \tilde{L},v_{12} S}^{\tilde{n}' \tilde{L}'}
\hat{V}_{\tilde{n} \tilde{L},v_{12} S}^{\tilde{n}' \tilde{L}'}
\label{eq:0},
\end{equation}
where the sum runs over the allowed indices of the tensors according
to the symmetries discussed below. Each term in the sum is
accompanied by the corresponding strength parameter
$C_{\tilde{n} \tilde{L},v_{12} S}^{\tilde{n}' \tilde{L}'}$, and explicitly reads,
\begin{eqnarray}
 \nonumber \hat{V}_{\tilde{n} \tilde{L},v_{12} S}^{\tilde{n}' \tilde{L}'}=
&&\frac{1}{2}i^{v_{12}} \left( \left[ \left[K'_{\tilde{n}'\tilde{L}'}
                                            K_ {\tilde{n }\tilde{L}}\right]_{S}
\SSS_{v_{12} S}\right]_{0}  \right. \\ \nonumber
&&\left. + (-1)^{v_{12}+S} \left[ \left[K'_{\tilde{n} \tilde{L}}
                                        K_ {\tilde{n}'\tilde{L}'}\right]_{S}
\SSS_{v_{12} S}\right]_{0}   \right) \\
&&\times \left(1-\hat{P}^{M}\hat{P}^{\sigma}\hat{P}^{\tau}\right)
\hat{\delta}_{12}(\bm{r}'_1\bm{r}'_2;\bm{r}_1\bm{r}_2)
\label{eq:1}.
\end{eqnarray}
In Eq.~(\ref{eq:1}), $K_ {\tilde{n }\tilde{L}}$ are
the spherical tensor derivatives of order $\tilde{n}$ and rank $\tilde{L}$
built of the spherical representations of the relative momenta
$\bm{k}=(\bm{\nabla}_1-\bm{\nabla}_2)/2i$,
\begin{eqnarray}
\label{eq:05a}
k_{1,\mu=\left\{-1,0,1\right\}} &=&
         -i       \left\{\tfrac{ 1}{\sqrt{2}}\left(k_x-ik_y\right), \right. \nonumber \\
                                &&  \left.         k_z,
                         \tfrac{-1}{\sqrt{2}}\left(k_x+ik_y\right)\right\} ;
\end{eqnarray}
up to sixth order they are listed in Table~\ref{tab:tb1}.
Similarly, operators  $K'_{\tilde{n }\tilde{L}}$ are built
of the relative momenta $\bm{k}'=(\bm{\nabla}'_1-\bm{\nabla}'_2)/2i$.

\begin{table}
\caption[T]{\label{tab:tb1} Derivative operators $K_{nL}$ up to N$^3$LO
as expressed through spherical tensor representation of
relative momenta $k$ defined in Eq.~(\ref{eq:05a}).}
\begin{center}\begin{tabular}{cccc}
\hline
\hline
No. & tensor  $K_{nL} $& order $n$ & rank $L$ \\
\hline
 1&$1 $& 0& 0 \\
 2&${k} $& 1& 1 \\
 3&$\lefts{k}{k}\rights_{0} $& 2& 0 \\
 4&$\lefts{k}{k}\rights_{2} $& 2& 2 \\
 5&$ \lefts{k}{k}\rights_{0} {k} $& 3& 1 \\
 6&$\lefts{k}\lefts{k}{k}\rights_{2}\rights_{3} $& 3& 3 \\
 7&$\lefts{k}{k}\rights_{0} ^{2} $& 4& 0 \\
 8&$\lefts{k}{k}\rights_{0} \lefts{k}{k}\rights_{2} $& 4& 2 \\
 9&$\lefts{k}\lefts{k}\lefts{k}{k}\rights_{2}\rights_{3}\rights_{4} $& 4& 4 \\
10&$\lefts{k}{k}\rights_{0}^2 {k} $& 5& 1 \\
11&$\lefts{k}{k}\rights_{0} \lefts{k}\lefts{k}{k}\rights_{2}\rights_{3} $& 5& 3 \\
12&$\lefts{k}\lefts{k}\lefts{k}\lefts{k}{k}\rights_{2}\rights_{3}\rights_{4}\rights_{5} $& 5& 5 \\
13&$\lefts{k}{k}\rights_{0} ^3 $& 6& 0 \\
14&$\lefts{k}{k}\rights_{0}^2 \lefts{k}{k}\rights_{2} $& 6& 2 \\
15&$\lefts{k}{k}\rights_{0} \lefts{k}\lefts{k}\lefts{k}{k}\rights_{2}\rights_{3}\rights_{4} $& 6& 4 \\
16&$\lefts{k}\lefts{k}\lefts{k}\lefts{k}\lefts{k}{k}\rights_{2}\rights_{3}\rights_{4}\rights_{5}\rights_{6} $& 6& 6 \\
\hline
\hline
\end{tabular}\end{center}
\end{table}

The symmetrized two-body spin operators $\SSS_{v_{12} S}$ are defined as,
\begin{eqnarray}
\label{eq:spin}
\SSS_{v_{12} S} =\left(1-\thalf\delta_{v_1,v_2}\right)\left(
[\sigma^{(1)}_{v_1}\sigma^{(2)}_{v_2}]_S +
[\sigma^{(1)}_{v_2}\sigma^{(2)}_{v_1}]_S \right),
\end{eqnarray}
where $v_{12}=v_1+v_2$ and $\sigma^{(i)}_{v\mu}$ are the spherical-tensor
components of the rank-$v$ Pauli matrices acting
on spin coordinates of particles $i=1$ or 2. They are expressed as
\begin{eqnarray}
\label{eq:03a}
\sigma^{(i)}_{00} &=& \hat{1} , \\
\label{eq:08a}
\sigma^{(i)}_{ 1,\mu=\left\{-1,0,1\right\}} &=&
     -i \left\{\tfrac{ 1}{\sqrt{2}}\left(\sigma^{(i)}_{ x}
                                       -i\sigma^{(i)}_{ y}\right), \right. \nonumber \\
                   && \left.             \sigma^{(i)}_{ z},
               \tfrac{-1}{\sqrt{2}}\left(\sigma^{(i)}_{ x}
                                       +i\sigma^{(i)}_{ y}\right)\right\}
\end{eqnarray}
through the spin unity matrix $\hat{1}$ and the standard Cartesian
components of the Pauli matrices $\sigma^{(i)}_{ x,y,z}$.

The Dirac delta function,
\begin{eqnarray}
\label{unit-pos}
\hspace*{-1.5em}
\hat{\delta}_{12}(\bbox{r}'_1\bbox{r}'_2,\bbox{r}_1\bbox{r}_2)
&=& \delta(\bbox{r}'_1\!-\!\bbox{r}_1)
\delta(\bbox{r}'_2\!-\!\bbox{r}_2)
\delta(\bbox{r} _1\!-\!\bbox{r}_2)
\nonumber \\
&=& \delta(\bbox{r}'_1\!-\!\bbox{r}_2)
\delta(\bbox{r}'_2\!-\!\bbox{r}_1)
\delta(\bbox{r} _2\!-\!\bbox{r}_1) .
\end{eqnarray}
ensures the locality and zero-range character of the pseudopotential.
The action of derivatives $K_ {\tilde{n }\tilde{L}}$ and  $K'_
{\tilde{n }\tilde{L}}$ on
$\hat{\delta}_{12}(\bbox{r}'_1\bbox{r}'_2,\bbox{r}_1\bbox{r}_2)$ has
to be understood in the standard sense of derivatives of
distributions. Whenever the pseudopotential (\ref{eq:0}) is inserted
into integrals to calculate the two-body matrix elements, the
integration by parts transfers the derivatives onto appropriate wave
functions in the remaining parts of integrands.

The exchange term is explicitly embedded in
the pseudopotential through the operator
\begin{eqnarray}
\nonumber \hat{P}^{M}\hat{P}^{\sigma}\hat{P}^{\tau}
&=&(-1)^{\tilde{n}'}\frac{1}{4}\bigg(1+\sqrt{3}\left[\sigma^{(1)}_1\sigma^{(2)}_1 \right]_0    \\
&&\hspace*{-1.4cm} + \sqrt{3}\left[\tau^{(1)}_1\tau^{(2)}_1 \right]^0
  + 3 \left[\sigma^{(1)}_1\sigma^{(2)}_1 \right]_0  \left[\tau^{(1)}_1\tau^{(2)}_1 \right]^0  \bigg)
\label{eq:2},
\end{eqnarray}
where $\tau^{(i)}_1$ are the standard spherical-tensor isospin Pauli
matrices defined analogously as in Eq.~(\ref{eq:08a}). The square
brackets with superscripts and subscripts denote the coupling of
spherical tensors in the isospin space and coordinate space,
respectively. The above definitions and conventions exactly
correspond to those introduced in Ref.~\cite{Carlsson1}.

The zero range of the pseudopotential has an important bearing on the
structure of terms in Eq.~(\ref{eq:1}). Indeed, only for the
zero-range force, the space-exchange (Majorana) operator
$\hat{P}^{M}$ can be replaced, in any individual term, by the phase
$(-1)^{\tilde{n}'}$ appearing in Eq.~(\ref{eq:2}). Moreover, apart
from the isospin-exchange operator $\hat{P}^{\tau}$, terms of the
pseudopotential cannot then depend on isospin. This fact, effectively
reduces by half the number of allowed terms of the pseudopotential,
as compared to what would have been possible for a finite-range
potential. This is at the origin of the numbers of allowed terms
of the pseudopotential being equal one half of the numbers of the
allowed terms of the EDF, which we discuss below.

The full antisymmetrization of the pseudopotential includes the
exchange operator in the isospin space; therefore, in the following
we consider the EDF with the isospin degree of freedom included, that
is, we discuss both the isoscalar and isovector terms of the N$^3$LO
\cite{Carlsson1}, which allows us to fully incorporate the
proton-neutron mixing at the level of the energy density
\cite{Perlinska1}.

The general form of the pseudopotential and the allowed terms listed
below reflect the fact that the fundamental symmetries of the
two-body interaction must be respected, see Appendix~\ref{appA}. In
particular, (i) all terms are scalar operators, that is, they are
coupled to the total angular momentum 0, which ensures the rotational
invariance, (ii) the total number of derivative operators must be
even, namely, $\tilde{n}+\tilde{n}'=0,2,4,6$, which ensures the
time-reversal and parity invariances, (iii) the parameters
$C_{\tilde{n} \tilde{L},v_{12} S}^{\tilde{n}' \tilde{L}'}$ of the
pseudopotential must be real, to guarantee both the time-reversal
invariance and hermiticity, and (iv) the invariance under exchange of
the coordinates of particle 1 and 2 is respected by expression
(\ref{eq:1}).

\subsection{\label{list pseudo}Lists of terms of the pseudopotential $\hat{V}$ order by order}

In Tables \ref{zero}-\ref{sixtable} are listed, respectively, all
possible terms of the pseudopotential (\ref{eq:0}) in zero,
second, fourth, and sixth order. In each order, the numbers of terms
equal 2, 7, 15, and 26, giving the total number of 50 terms up to
N$^3$LO. We see that these numbers of terms are exactly equal to
those corresponding to the EDF in {\em each} isospin channel with the
Galilean invariance imposed, cf.~Table~VI of Ref.~\cite{Carlsson1}.
One should note that each term of the pseudopotential (\ref{eq:1}) is
Galilean-invariant by construction, because it is built with
relative-momentum operators $K_{\tilde{n}\tilde{L}}$; therefore, the
pseudopotential is not changed by a transformation to a system moving
with a constant velocity. When both isoscalar and isovector channels
are considered in the EDF, the number of EDF terms becomes in each
order {\em twice larger} than the number of terms of the
pseudopotential.

This means that the EDF obtained by averaging the pseudopotential is
constrained by as many conditions as there are terms in each isospin
channel. One possible solution is than to find a one-to-one
correspondence between the EDF and the pseudopotential by relating
the isoscalar part of the EDF to its isovector part, in a way that
will be showed explicitly in the following Sections of this work.

\begin{table}
\caption{\label{zero}Zero-order terms of the pseudopotential (\ref{eq:1}).}
\begin{ruledtabular}
\begin{tabular}{cccccccc}
No.     &  $\tilde{n}'$ &  $\tilde{L}'$ &  $\tilde{n}$  &  $\tilde{L}$  &    $v_{12}$   &     $S$        &    gauge      \\
\hline
1       &       0       &       0       &       0       &       0       &       0       &       0        &       Y       \\
2       &       0       &       0       &       0       &       0       &       2       &       0        &       Y       \\
\end{tabular}
\end{ruledtabular}
\end{table}

\begin{table}
\caption{\label{sectable}Same as in Table \ref{zero} but for the second order terms.}
\begin{ruledtabular}
\begin{tabular}{cccccccc}
No.     &  $\tilde{n}'$ &  $\tilde{L}'$ &  $\tilde{n}$  &  $\tilde{L}$  &    $v_{12}$   &     $S$       &    gauge      \\
\hline
1       &       2       &       0       &       0       &       0       &       0       &       0       &       Y       \\
2       &       2       &       0       &       0       &       0       &       2       &       0       &       Y       \\
3       &       2       &       2       &       0       &       0       &       2       &       2       &       Y       \\
4       &       1       &       1       &       1       &       1       &       0       &       0       &       Y       \\
5       &       1       &       1       &       1       &       1       &       2       &       0       &       Y       \\
6       &       1       &       1       &       1       &       1       &       1       &       1       &       Y       \\
7       &       1       &       1       &       1       &       1       &       2       &       2       &       Y       \\
\end{tabular}
\end{ruledtabular}
\end{table}

\begin{table}[!ht]
\caption{\label{fourtable}Same as in Table \ref{zero} but for the fourth order terms.}
\begin{ruledtabular}
\begin{tabular}{cccccccc}
No.     &  $\tilde{n}'$ &  $\tilde{L}'$ &  $\tilde{n}$  &  $\tilde{L}$  &    $v_{12}$   &     $S$       &    gauge      \\
\hline
 1      &       4       &       0       &       0       &       0       &       0       &       0       &       D       \\
 2      &       4       &       0       &       0       &       0       &       2       &       0       &       D       \\
 3      &       4       &       2       &       0       &       0       &       2       &       2       &       D       \\
 4      &       3       &       1       &       1       &       1       &       0       &       0       &       Y       \\
 5      &       3       &       1       &       1       &       1       &       2       &       0       &       Y       \\
 6      &       3       &       1       &       1       &       1       &       1       &       1       &       N       \\
 7      &       3       &       1       &       1       &       1       &       2       &       2       &       D       \\
 8      &       3       &       3       &       1       &       1       &       2       &       2       &       I       \\
 9      &       2       &       0       &       2       &       0       &       0       &       0       &       D       \\
 10     &       2       &       0       &       2       &       0       &       2       &       0       &       D       \\
 11     &       2       &       2       &       2       &       0       &       2       &       2       &       D       \\
 12     &       2       &       2       &       2       &       2       &       0       &       0       &       I       \\
 13     &       2       &       2       &       2       &       2       &       2       &       0       &       I       \\
 14     &       2       &       2       &       2       &       2       &       1       &       1       &       N       \\
 15     &       2       &       2       &       2       &       2       &       2       &       2       &       I       \\
\end{tabular}
\end{ruledtabular}
\end{table}

\begin{table}
\caption{\label{sixtable}Same as in Table \ref{zero} but for the sixth order terms.}
\begin{ruledtabular}
\begin{tabular}{cccccccc}
No.     &  $\tilde{n}'$ &  $\tilde{L}'$ &  $\tilde{n}$  &  $\tilde{L}$  &    $v_{12}$   &     $S$       &    gauge      \\
\hline
 1      &       6       &       0       &       0       &       0       &       0       &       0       &       D       \\
 2      &       6       &       0       &       0       &       0       &       2       &       0       &       D       \\
 3      &       6       &       2       &       0       &       0       &       2       &       2       &       D       \\
 4      &       5       &       1       &       1       &       1       &       0       &       0       &       D       \\
 5      &       5       &       1       &       1       &       1       &       2       &       0       &       D       \\
 6      &       5       &       1       &       1       &       1       &       1       &       1       &       N       \\
 7      &       5       &       1       &       1       &       1       &       2       &       2       &       D       \\
 8      &       5       &       3       &       1       &       1       &       2       &       2       &       I       \\
 9      &       4       &       0       &       2       &       0       &       0       &       0       &       D       \\
 10     &       4       &       0       &       2       &       0       &       2       &       0       &       D       \\
 11     &       4       &       2       &       2       &       0       &       2       &       2       &       D       \\
 12     &       4       &       0       &       2       &       2       &       2       &       2       &       D       \\
 13     &       4       &       2       &       2       &       2       &       0       &       0       &       I       \\
 14     &       4       &       2       &       2       &       2       &       2       &       0       &       I       \\
 15     &       4       &       2       &       2       &       2       &       1       &       1       &       N       \\
 16     &       4       &       2       &       2       &       2       &       2       &       2       &       D       \\
 17     &       4       &       4       &       2       &       2       &       2       &       2       &       I       \\
 18     &       3       &       1       &       3       &       1       &       0       &       0       &       D       \\
 19     &       3       &       1       &       3       &       1       &       2       &       0       &       D       \\
 20     &       3       &       1       &       3       &       1       &       1       &       1       &       N       \\
 21     &       3       &       1       &       3       &       1       &       2       &       2       &       D       \\
 22     &       3       &       3       &       3       &       1       &       2       &       2       &       D       \\
 23     &       3       &       3       &       3       &       3       &       0       &       0       &       I       \\
 24     &       3       &       3       &       3       &       3       &       2       &       0       &       I       \\
 25     &       3       &       3       &       3       &       3       &       1       &       1       &       N       \\
 26     &       3       &       3       &       3       &       3       &       2       &       2       &       D       \\
\end{tabular}
\end{ruledtabular}
\end{table}

To make the connection between the pseudopotential and the
standard form of the Skyrme interaction more transparent, we give
here the relations of conversion between the parameters of the zero- and second-order
pseudopotential and those of the Skyrme
interaction, see Ref.~\cite{Perlinska1} for the definitions used.
They read,
\begin{subequations}
\label{eq:conversion}
\begin{eqnarray}
t_0   &=&C_{00,00}^{00}+\frac{1}{\sqrt{3}}C_{00,20}^{00}, \label{subeq:0conversion} \\
t_0x_0&=&-\frac{2}{\sqrt{3}}C_{00,20}^{00}, \label{subeq:0bisconversion} \\
t_1   &=& \frac{1}{\sqrt{3}}C_{00,00}^{20}+\frac{1}{3}C_{00,20}^{20}, \label{subeq:1conversion} \\
t_1x_1&=& -\frac{2}{3}C_{00,20}^{20}, \label{subeq:2conversion} \\
t_2   &=& \frac{1}{\sqrt{3}}C_{11,00}^{11}+\frac{1}{3}C_{11,20}^{11}, \label{subeq:3conversion}  \\
t_2x_2&=& -\frac{2}{3}C_{11,20}^{11}, \label{subeq:4conversion}  \\
W_0   &=&\frac{1}{\sqrt{6}}C_{11,11}^{11}, \label{subeq:5conversion}  \\
t_o   &=&-\frac{1}{3\sqrt{5}}C_{11,22}^{11}, \label{subeq:6conversion}  \\
t_e   &=&-\frac{1}{3\sqrt{5}}C_{00,22}^{22}. \label{subeq:7conversion}
\end{eqnarray}
\end{subequations}

In relations of Eqs.~(\ref{eq:conversion}), parameters $t_3$ and
$t_3x_3$ are missing: they are related to the terms of the Skyrme
interaction depending on density, which have been introduced to mimic
the effects of the three-body force in the phenomenological
interaction and to get the saturation feature of the nuclear force.
In the same way, the zero-order parameters $C_{00,00}^{00}$ and
$C_{00,20}^{00}$ of the pseudopotential, see
Eqs.~(\ref{subeq:0conversion}) and  (\ref{subeq:0bisconversion}),
should become density-dependent.

In his effective nuclear potential, Skyrme also introduced~\cite{Skyrme2} one
additional term of the fourth order, which he justified through the
presence of considerable D-waves in the nucleon-nucleon interaction
energies around 100\,MeV. Also in this case, we give the relation
between the corresponding parameter $t_{D}$ and the parameter of
our full pseudopotential,

\begin{equation}
t_D   =\frac{1}{2}C_{20,20}^{00}
\label{eq:conversionfourth}.
\end{equation}

\subsection{\label{subsec:Gauge}Gauge invariance of the pseudopotential}

Besides the Galilean invariance mentioned above, the standard Skyrme force
has been also proved to be invariant with respect to a more general
local gauge invariance, and to give rise to the energy density that is
invariant under the same symmetry when specific relations between
the coupling constants are set \cite{Engel1,Dobaczewski2}.

The gauge transformation acts on a many-body wave function by
multiplying it with a position-dependent phase factor, that is,
\begin{equation}
|\Psi'\rangle = \exp \left( i \sum_{j=1}^{A}\phi(r_j) \right)|\Psi\rangle
\label{eq:3},
\end{equation}
and its action transferred onto the pseudopotential is,
\begin{equation}
\hat{V'}= e^{-i\phi(r'_2)} e^{-i\phi(r'_1)} \hat{V} e^{i\phi(r_1)} e^{i\phi(r_2)}
\label{eq:4}.
\end{equation}

Apart from zero order, the terms of the pseudopotential are not trivially invariant with
respect to the transformation of the Eq.~(\ref{eq:4}) and, in general,
the transformed pseudopotential $\hat{V}'$ is different than the
original pseudopotential $\hat{V}$. To impose the gauge invariance on
the pseudopotential, one has to derive a list of constraints among the
parameters, which can be done using the condition
\begin{equation}
[\phi(r_1),\hat{V}] + [\phi(r_2),\hat{V}] = 0 .
\label{eq:5}
\end{equation}
As expected, at second order, all the 7 terms of the pseudopotential listed in
Table~\ref{sectable}
fulfill condition~(\ref{eq:5}).
Then they all are the stand-alone gauge invariant terms of the
pseudopotential, which in the last column of the Table is marked by the
letter Y. On the other hand, at fourth order, only two of the
terms of the pseudopotential listed in Table~\ref{fourtable},
those that correspond to parameters $C_{11,00}^{31}$ and $C_{11,20}^{31}$,
fulfill condition~(\ref{eq:5}). At sixth order, none of the terms
are stand-alone gauge invariant.

At fourth order, the gauge invariance forces seven parameters of the
pseudopotential to be specific linear combinations of four
independent ones. In Table~\ref{fourtable}, they are marked by
letters D and I, respectively. In Appendix~\ref{appB}, we list such
relations between the dependent and independent parameters. One
should note that other choices of the four independent parameters are
also possible, that is, at fourth order, there are simply four
different gauge-invariant linear combinations of terms of the
pseudopotential (\ref{eq:0}). Moreover, at this order, there are also
two terms that alone are gauge non-invariant -- those that correspond
to parameters $C_{11,11}^{31}$ and $C_{22,11}^{22}$; in
Table~\ref{fourtable}, they are marked by letters N. Similarly, at
sixth order, there are six gauge-invariant linear combinations of
terms of the pseudopotential, that is, sixteen dependent parameters
are related to six independent ones, see Appendix~\ref{appB}, and
there are also four alone gauge non-invariant terms corresponding to
parameters $C_{11,11}^{51}$, $C_{22,11}^{42}$, $C_{31,11}^{31}$, and
$C_{33,11}^{33}$.

\begin{figure}
\includegraphics[width=7cm]{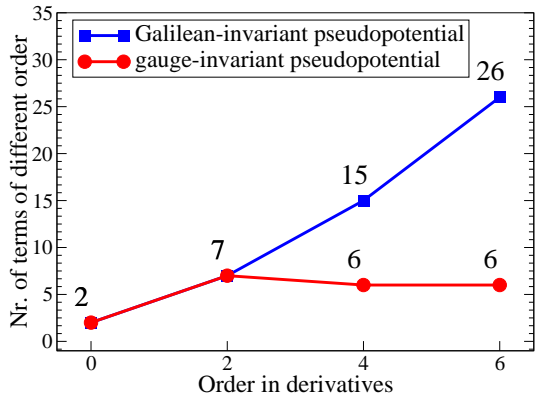}
\caption{\label{numterm}(Color online) Number of terms of the pseudopotential
(\ref{eq:1}), plotted as a function of the order in derivatives.}
\end{figure}

A comparison between the numbers of terms of the Galilean-invariant
pseudopotential and the gauge-invariant pseudopotential is plotted in
Fig.~\ref{numterm}. Again we note that at each order, the numbers of
gauge-invariant parameters (2 for the zero order, 7 for the second
order, 6 for the fourth order, and 6 for the sixth order) are exactly
the same as the numbers of independent coupling constants of the EDF
in {\em each} isospin channel with the gauge invariance imposed,
cf.~Table~VI of Ref.~\cite{Carlsson1}. Again, this observation will
be crucial when we proceed to derive relations between the isoscalar
and the isovector parts of the EDF, stemming from the gauge-invariant
pseudopotential. We also remark that whereas the second-order
spin-orbit term, corresponding to parameter $C_{11,11}^{11}$, is
gauge invariant, all higher-order spin-orbit terms, corresponding to
parameters $C_{\tilde{n} \tilde{L},11}^{\tilde{n}' \tilde{L}'}$ with
$\tilde{n}+\tilde{n}'>2$ do violate the gauge symmetry.

\subsection{\label{subsec:tensor-like pseudopotential} Tensor-like form of the pseudopotential}
In this Section, we present the tensor-like form of the
pseudopotential, which is, in fact, a different form of coupling of
the relative-momentum operators with the spin operators, just like in
the tensor term of the standard Skyrme interaction. In this form, the pseudopotential
of Eq.~(\ref{eq:0}) is a sum of the following terms,
\begin{equation}
 \hat{V}=\sum_{\substack{\tilde{n}' \tilde{L}', \\ \tilde{n} \tilde{L},v_{12} J}}
      \tilde{C}_{\tilde{n} \tilde{L},v_{12} J}^{\tilde{n}' \tilde{L}'}
\hat{\tilde{V}}_{\tilde{n} \tilde{L},v_{12} J}^{\tilde{n}' \tilde{L}'} ,
\label{eq:0t}
\end{equation}
where
\begin{eqnarray}
\nonumber
\hat{\tilde{V}}_{\tilde{n} \tilde{L}, v_{12}J}^{\tilde{n}' \tilde{L}'}
=&&\frac{1}{2} i^{v_{12}} \left(1-\thalf\delta_{v_1,v_2}\right) \times \\ \nonumber
&& \left( \left[ \left[K'_{\tilde{n}'\tilde{L}'} \sigma^{(1)}_{v_1} \right]_{J}
       \left[K_{\tilde{n}\tilde{L}} \sigma^{(2)}_{v_2} \right]_{J}\right]_{0}  \right. \\ \nonumber
&&\left. + \left[ \left[K'_{\tilde{n}'\tilde{L}'} \sigma^{(2)}_{v_1} \right]_{J}
       \left[K_{\tilde{n}\tilde{L}} \sigma^{(1)}_{v_2} \right]_{J}\right]_{0}  \right. \\ \nonumber
&&\left. + \left[ \left[K'_{\tilde{n}\tilde{L}} \sigma^{(1)}_{v_1} \right]_{J}
           \left[K_{\tilde{n}'\tilde{L}'} \sigma^{(2)}_{v_2} \right]_{J}\right]_{0} \right. \\ \nonumber
&&\left. + \left[ \left[K'_{\tilde{n}\tilde{L}} \sigma^{(2)}_{v_1} \right]_{J}
           \left[K_{\tilde{n}'\tilde{L}'} \sigma^{(1)}_{v_2} \right]_{J}\right]_{0}
\right)\times \\
&&\left(1-\hat{P}^{M}\hat{P}^{\sigma}\hat{P}^{\tau}\right)
\hat{\delta}_{12}(\bm{r}'_1\bm{r}'_2;\bm{r}_1\bm{r}_2)
\label{eq:9}.
\end{eqnarray}

The lists of the zero-, second-, fourth-, and sixth-order terms
$\hat{\tilde{V}}_{\tilde{n} \tilde{L},v_{12} J}^{\tilde{n}'
\tilde{L}'}$ of the pseudopotential are given, respectively, in
Tables~\ref{zerotablerec}--\ref{sixtablerec},
which are the analogues of Tables~\ref{zero}--\ref{sixtable} given
in Section~\ref{list pseudo}.

\begin{table}
\caption{\label{zerotablerec}Zero-order terms of the recoupled pseudopotential ($\ref{eq:9}$).}
\begin{ruledtabular}
\begin{tabular}{ccccccc}
No.     &  $\tilde{n}'$ &  $\tilde{L}'$ &  $\tilde{n}$  &  $\tilde{L}$  &    $v_{12}$   &     $J$       \\
\hline
1       &       0       &       0       &       0       &       0       &       0       &       0       \\
2       &       0       &       0       &       0       &       0       &       2       &       0       \\
\end{tabular}
\end{ruledtabular}
\end{table}

\begin{table}
\caption{\label{sectablerec}Same as in Table \ref{zerotablerec} but for the second-order terms.}
\begin{ruledtabular}
\begin{tabular}{ccccccc}
No.     &  $\tilde{n}'$ &  $\tilde{L}'$ &  $\tilde{n}$  &  $\tilde{L}$  &    $v_{12}$   &     $J$       \\
\hline
1       &       1       &       1       &       1       &       1       &       0       &       1       \\
2       &       1       &       1       &       1       &       1       &       1       &       1       \\
3       &       1       &       1       &       1       &       1       &       2       &       0       \\
4       &       1       &       1       &       1       &       1       &       2       &       1       \\
5       &       1       &       1       &       1       &       1       &       2       &       2       \\
6       &       2       &       0       &       0       &       0       &       0       &       0       \\
7       &       2       &       0       &       0       &       0       &       2       &       1       \\
8       &       2       &       2       &       0       &       0       &       2       &       1       \\
\end{tabular}
\end{ruledtabular}
\end{table}

\begin{table}
\caption{\label{fourtablerec}Same as in Table \ref{zerotablerec} but for the fourth-order terms.}
\begin{ruledtabular}
\begin{tabular}{ccccccc}
No.     &  $\tilde{n}'$ &  $\tilde{L}'$ &  $\tilde{n}$  &  $\tilde{L}$  &    $v_{12}$   &     $J$       \\
\hline
 1      &       2       &       0       &       2       &       0       &       0       &       0       \\
 2      &       2       &       0       &       2       &       0       &       2       &       1       \\
 3      &       2       &       2       &       2       &       2       &       0       &       2       \\
 4      &       2       &       2       &       2       &       2       &       1       &       2       \\
 5      &       2       &       2       &       2       &       0       &       2       &       1       \\
 6      &       2       &       2       &       2       &       2       &       2       &       1       \\
 7      &       2       &       2       &       2       &       2       &       2       &       2       \\
 8      &       2       &       2       &       2       &       2       &       2       &       3       \\
 9      &       3       &       1       &       1       &       1       &       0       &       1       \\
 10     &       3       &       1       &       1       &       1       &       1       &       1       \\
 11     &       3       &       1       &       1       &       1       &       2       &       0       \\
 12     &       3       &       1       &       1       &       1       &       2       &       1       \\
 13     &       3       &       1       &       1       &       1       &       2       &       2       \\
 14     &       3       &       3       &       1       &       1       &       2       &       2       \\
 15     &       4       &       0       &       0       &       0       &       0       &       0       \\
 16     &       4       &       0       &       0       &       0       &       2       &       1       \\
 17     &       4       &       2       &       0       &       0       &       2       &       1       \\
 \\
\end{tabular}
\end{ruledtabular}
\end{table}

\begin{table}
\caption{\label{sixtablerec}Same as in Table \ref{zerotablerec} but for the sixth-order terms.}
\begin{ruledtabular}
\begin{tabular}{ccccccc}
No.     &  $\tilde{n}'$ &  $\tilde{L}'$ &  $\tilde{n}$  &  $\tilde{L}$  &    $v_{12}$   &     $J$       \\
\hline
 1      &       3       &       1       &       3       &       1       &       0       &       1       \\
 2      &       3       &       1       &       3       &       1       &       1       &       1       \\
 3      &       3       &       1       &       3       &       1       &       2       &       0       \\
 4      &       3       &       1       &       3       &       1       &       2       &       1       \\
 5      &       3       &       1       &       3       &       1       &       2       &       2       \\
 6      &       3       &       3       &       3       &       3       &       0       &       3       \\
 7      &       3       &       3       &       3       &       3       &       1       &       3       \\
 8      &       3       &       3       &       3       &       1       &       2       &       2       \\
 9      &       3       &       3       &       3       &       3       &       2       &       2       \\
 10     &       3       &       3       &       3       &       3       &       2       &       3       \\
 11     &       3       &       3       &       3       &       3       &       2       &       4       \\
 12     &       4       &       0       &       2       &       0       &       0       &       0       \\
 13     &       4       &       0       &       2       &       0       &       2       &       1       \\
 14     &       4       &       0       &       2       &       2       &       2       &       1       \\
 15     &       4       &       2       &       2       &       2       &       0       &       2       \\
 16     &       4       &       2       &       2       &       2       &       1       &       2       \\
 17     &       4       &       2       &       2       &       0       &       2       &       1       \\
 18     &       4       &       2       &       2       &       2       &       2       &       1       \\
 19     &       4       &       2       &       2       &       2       &       2       &       2       \\
 20     &       4       &       2       &       2       &       2       &       2       &       3       \\
 21     &       4       &       4       &       2       &       2       &       2       &       3       \\
 22     &       5       &       1       &       1       &       1       &       0       &       1       \\
 23     &       5       &       1       &       1       &       1       &       1       &       1       \\
 24     &       5       &       1       &       1       &       1       &       2       &       0       \\
 25     &       5       &       1       &       1       &       1       &       2       &       1       \\
 26     &       5       &       1       &       1       &       1       &       2       &       2       \\
 27     &       5       &       3       &       1       &       1       &       2       &       2       \\
 28     &       6       &       0       &       0       &       0       &       0       &       0       \\
 29     &       6       &       0       &       0       &       0       &       2       &       1       \\
 30     &       6       &       2       &       0       &       0       &       2       &       1       \\
\end{tabular}
\end{ruledtabular}
\end{table}

By means of the recoupling technique, it is possible to determine
relations between the two different coupling schemes of the pseudopotential.
This derivation, along with the relationships between the corresponding
parameters
$C_{\tilde{n} \tilde{L},v_{12} S}^{\tilde{n}' \tilde{L}'}$ and
$\tilde{C}_{\tilde{n} \tilde{L},v_{12} J}^{\tilde{n}' \tilde{L}'}$,
is presented in Appendix~\ref{appD}.

The reader might have noticed that the two forms of the
pseudopotential do not have the same numbers of terms: the
tensor-like form of the pseudopotential (Tables~\ref{sectablerec},
\ref{fourtablerec}, and \ref{sixtablerec}) has more terms than the
central-like form (Tables~\ref{sectable}, \ref{fourtable}, and
\ref{sixtable}). This means that not all of the terms of the tensor-like
form are linearly independent from one another, even though they are
all allowed by the symmetries, and thus some terms can be expressed
as linear combinations of others, or, equivalently, some linear
combinations of terms are identically equal to zero. This fact, can
be expressed in the form of the following explicit dependencies
between the parameters of the tensor-like pseudopotential.

For the second-order terms we have,
\begin{equation}
\tilde{C}_{11,21}^{11}=-\frac{2}{\sqrt{3}}\tilde{C}_{11,20}^{11}+\sqrt{\frac{5}{3}}\tilde{C}_{11,22}^{11}
\label{eq:13},
\end{equation}
whereas the fourth-order dependencies read,
\begin{subequations}
\label{eq:14}
\begin{eqnarray}
\tilde{C}_{22,21}^{22}&=&-\frac{\sqrt{15}}{9}\tilde{C}_{22,22}^{22} + \frac{2}{9}\sqrt{21}\tilde{C}_{22,23}^{22}  ,  \label{subeq: 1of14}  \\
\tilde{C}_{11,21}^{31}&=&-\frac{2}{\sqrt{3}}\tilde{C}_{11,20}^{31}+\sqrt{\frac{5}{3}}\tilde{C}_{11,22}^{31} , \label{subeq: 2of14}
\end{eqnarray}
\end{subequations}
and finally at sixth order we have,
\begin{subequations}
\label{eq:15}
\begin{eqnarray}
\tilde{C}_{31,21}^{31}&=& -\frac{2}{\sqrt{3}}\tilde{C}_{31,20}^{31} + \sqrt{\frac{5}{3}}\tilde{C}_{31,22}^{31} , \label{subeq: 1of15}   \\
\tilde{C}_{33,23}^{33}&=& -4 \sqrt{\frac{5}{7}}\tilde{C}_{33,22}^{33}+\frac{9}{\sqrt{7}}\tilde{C}_{33,24}^{33} ,  \label{subeq: 2of15}  \\
\tilde{C}_{22,21}^{42}&=& -\frac{\sqrt{15}}{9}\tilde{C}_{22,22}^{42}+\frac{2}{9}\sqrt{21}\tilde{C}_{22,23}^{42} ,  \label{subeq: 3of15}  \\
\tilde{C}_{11,21}^{51}&=& -\frac{2}{\sqrt{3}}\tilde{C}_{11,20}^{51}+\sqrt{\frac{5}{3}}\tilde{C}_{11,22}^{51} .\label{subeq: 4of15}
\end{eqnarray}
\end{subequations}

\section{\label{sec:isoEDF}Relations between the pseudopotential and
Energy Density Functional}

The EDF related to the pseudopotential is obtained by averaging the
pseudopotential $\hat{V}$ over the uncorrelated wavefunction (a Slater
determinant), that is,
\begin{widetext}
\begin{eqnarray}
\label{eq:16}
{\cal E}&=&\frac{1}{4} \int {\rm d}\, \bm{r}_{1}\bm{r}_{2}\bm{r}'_{1}\bm{r}'_{2}
\sum_{\substack{{s_1}{s_2} \\{s'_1}{s'_2}}}
\sum_{\substack{{t_1}{t_2} \\{t'_1}{t'_2}}}
\hat{V}(\bm{r}'_{1} s'_1 t'_1 \bm{r}'_{2} s'_2 t'_2, \bm{r}_{1} s_1 t_1 \bm{r}_{2} s_2 t_2)
\rho(\bm{r}_{1} s_1 t_1,\bm{r}'_{1} s'_1 t'_1)
\rho(\bm{r}_{2} s_2 t_2,\bm{r}'_{2} s'_2 t'_2),
\end{eqnarray}
where the two-body spin-isospin matrix element of the pseudopotential is defined as
\begin{eqnarray}
\label{eq:16a}
\hat{V}(\bm{r}'_{1} s'_1 t'_1 \bm{r}'_{2} s'_2 t'_2, \bm{r}_{1} s_1 t_1 \bm{r}_{2} s_2 t_2)
=\langle s'_1 t'_1, s'_2 t'_2 |\hat{V}|s_1 t_1 , s_2 t_2\rangle  ,
\end{eqnarray}
\end{widetext}
and
$\rho(\bm{r}_{1} s_1 t_1,\bm{r}'_{1} s'_1 t'_1)$ and $\rho(\bm{r}_{2} s_2 t_2,\bm{r}'_{2} s'_2 t'_2)$,
are the one-body densities in spin-isospin channels. (For
definitions, see, e.g., Ref.~\cite{Perlinska1}.) In this lengthy
calculation, one must consider as intermediate step the recoupling of
the relative-momentum operators, so as to recast the gradients in
such a way that each tensor affects only one particle at a time
\cite{Perlinska1}. Such recoupling was performed with the aid of
symbolic programming, and is not, for the sake of brevity, reported
in this paper.

For each term of the pseudopotential (\ref{eq:0}), we can write the
result of the averaging in the following way,
\begin{equation}
\langle C_{\tilde{n} \tilde{L},v_{12} S}^{\tilde{n}' \tilde{L}'}
\hat{V}_{\tilde{n} \tilde{L},v_{12} S}^{\tilde{n}' \tilde{L}'} \rangle =
 \sum C_{m I,n L v J}^{n' L' v' J',t} T_{m I, n L v J}^{n' L' v' J', t} \label{eq:17},
\end{equation}
where $C_{m I,n L v J}^{n' L' v' J', t}$ and $T_{m I, n L v J}^{n' L' v' J', t}$
denote, respectively, the coupling constants and terms of the EDF
according to the formalism developed in Ref.~\cite{Carlsson1}. Since
here we treat the isospin degree of freedom explicitly, to the
notation of Ref.~\cite{Carlsson1} we have added superscripts $t$,
which denote the isoscalar ($t=0$) and isovector ($t=1$) channels.

Once relations (\ref{eq:17}) are evaluated for each term of the
pseudopotential, all terms of the N$^3$LO EDF are generated, with the
EDF coupling constants $C_{m I,n L v J}^{n' L' v' J', t}$ becoming
linear combinations of the pseudopotential strength parameters
$C_{\tilde{n} \tilde{L},v_{12} S}^{\tilde{n}' \tilde{L}'}$. Since the
pseudopotentials are Galilean-invariant, the obtained EDF coupling
constants obey the Galilean-invariance constraints~\cite{Carlsson1}.
Similarly, when parameters of the pseudopotential are restricted
to obey the gauge-invariance conditions defined in
Sec.~\ref{subsec:Gauge}, the resulting coupling constants correspond
to a gauge-invariant EDF.

The 12 second-order isoscalar (isovector) coupling constants
expressed by the 7 second-order pseudopotential parameters are given
in Table \ref{second energy isoscalar} (Table \ref{second energy
isovector}). Similar expressions relating at fourth (sixth) order 45
(129) isoscalar and isovector coupling constants to 15 (26)
pseudopotential parameters, are available in the supplemental
material~\cite{suppl}.

\begin{table}
\caption{\label{second energy isoscalar}Second-order coupling
constants of the isoscalar EDF $(t=0)$ as functions of parameters of the
pseudopotential, expressed by the formula $C_{mI,n L v J}^{n' L' v'
J', 0}= A(a C_{00,00}^{20}+b C_{00,20}^{20}+c C_{00,22}^{22}+d
C_{11,00}^{11}+e C_{11,20}^{11}+f C_{11,11}^{11}+g C_{11,22}^{11})$.}
\begin{ruledtabular}
\begin{tabular}{ccccccccc}
& $A$ & $a$ & $b$ & $c$ & $d$ & $e$ & $f$ &$g$ \\
\hline
$C_{20,0000}^{0000,0}$       &    $\frac{1}{32}$          &     $-$3       &      $-\sqrt{3}$        &       0       &      5        &       $-\sqrt{3}$         &   0      & 0  \\
$C_{00,2000}^{0000,0}$       &     $\frac{1}{16}$        &      3        &      $\sqrt{3}$         &       0       &         5     &      $-\sqrt{3}$         &     0    &  0 \\
$C_{00,1110}^{1110,0}$       &     $\frac{1}{48}$          &      $\sqrt{3}$        &     $5$         &      $2\sqrt{5}$         &        $-\sqrt{3}$       &   3      &      0    &   $6\sqrt{5}$  \\
$C_{00,1111}^{1111,0}$       &    $\frac{1}{48}$           &     3         &     $5\sqrt{3}$          &      $-\sqrt{15}$          &      $-$3      &   $3\sqrt{3}$           &   0    & $-3\sqrt{15}$   \\
$C_{00,1112}^{1112,0}$       &    $\frac{1}{48}$             &   $\sqrt{15}$   &    $5\sqrt{5}$       &       1   &     $-\sqrt{15}$    &    $3\sqrt{5}$    & 0 & 3    \\
$C_{11,1111}^{0000,0}$       &     $-\frac{3}{4}$           &    0          &    0          &        0      &       0       &      0        &    1   &  0\\
$C_{00,1101}^{1101,0}$       &     $\frac{1}{16}$           &    $-$3        &     $-\sqrt{3}$         &        0      &       $-$5     &      $\sqrt{3}$        &    0   & 0  \\
$C_{20,0011}^{0011,0}$       &      $\frac{1}{32}$          &     $\sqrt{3}$         &      5        &      0        &      $\sqrt{3}$        &    $-$3        &  0    &  0 \\
$C_{22,0011}^{0011,0}$       &      $\frac{1}{16}$           &     0         &     0         &      1        &       0       &     0         &    0   &$-$3 \\
$C_{00,2011}^{0011,0}$       &     $\frac{1}{16}$           &     $-\sqrt{3}$          &     $-$5       &     0         &      $\sqrt{3}$          &     $-$3       &  0     &  0 \\
$C_{00,2211}^{0011,0}$       &     $\frac{1}{8}$          &     0         &        0      &      $-$1      &      0        &    0          &  0     &$-$3 \\
$C_{11,0011}^{1101,0}$       &      $-\frac{3}{4}$          &     0         &        0      &       0       &         0     &     0        &   1    & 0 \\
\end{tabular}
\end{ruledtabular}
\end{table}

\begin{table}
\caption{\label{second energy isovector}Same as in Table \ref{second
energy isoscalar} but for isovector EDF $(t=1)$, according to the formula
$C_{mI,n L v J}^{n' L' v' J', 1}= A(a C_{00,00}^{20}+b
C_{00,20}^{20}+c C_{00,22}^{22}+d C_{11,00}^{11}+e C_{11,20}^{11}+f
C_{11,11}^{11}+g C_{11,22}^{11})$.}
\begin{ruledtabular}
\begin{tabular}{ccccccccc}
& $A$ & $a$ & $b$ & $c$ & $d$ & $e$ & $f$ &$g$ \\
\hline
$C_{20,0000}^{0000,1}$       &      $\frac{1}{32}$             &     $\sqrt{3}$         &    $-$3        &       0       &      $\sqrt{3}$         &   $-$3         &    0  &  0 \\
$C_{00,2000}^{0000,1}$       &        $\frac{1}{16}$           &      $-\sqrt{3}$        &        3      &      0        &      $\sqrt{3}$        &    $-$3        &    0  &  0 \\
$C_{00,1110}^{1110,1}$       &    $\frac{1}{48} $         &         3     &     $\sqrt{3}$         &       $-2\sqrt{15}$       &  $-$3          &    $-\sqrt{3}$           &    0   & $2\sqrt{15}$ \\
$C_{00,1111}^{1111,1}$       &      $\frac{1}{16} $           &     $\sqrt{3}$         &      $1$         &       $\sqrt{5}$         &      $-\sqrt{3}$         &      $-1$         &   0    & $-\sqrt{5}$  \\
$C_{00,1112}^{1112,1}$       &     $\frac{1}{48} $             &    $3\sqrt{5}$           &      $\sqrt{15}$          &     $-\sqrt{3}$          &    $-3\sqrt{5}$             &     $-\sqrt{15}$            &   0     & $\sqrt{3}$ \\
$C_{11,1111}^{0000,1}$       &      $-\frac{1}{4}\sqrt{3} $            &      0        &        0      &          0    &         0     &      0        &     1 &  0 \\
$C_{00,1101}^{1101,1}$       &    $\frac{1}{16}$          &      $\sqrt{3}$        &     $-$3       &         0     &        $-\sqrt{3}$      &    3          &  0     & 0 \\
$C_{20,0011}^{0011,1}$       &    $\frac{1}{32}$          &     3         &     $\sqrt{3}$         &        0      &        3      &      $\sqrt{3}$        &  0    & 0  \\
$C_{22,0011}^{0011,1}$       &      $-\frac{1}{16}\sqrt{3}$           &    0      &       0       &       1       &       0       &     0         &   0   &  1  \\
$C_{00,2011}^{0011,1}$       &      $\frac{1}{16}$          &    $-$3        &       $-\sqrt{3}$       &      0        &      3        &      $\sqrt{3}$        &    0  &  0 \\
$C_{00,2211}^{0011,1}$       &       $\frac{1}{8}\sqrt{3} $           &        0      &        0      &      1        &        0      &     0         &    0  &$-$1 \\
$C_{11,0011}^{1101,1}$       &      $-\frac{1}{4}\sqrt{3} $         &      0        &        0      &         0     &        0      &       0       &  1    &   0 \\
\end{tabular}
\end{ruledtabular}
\end{table}

\subsection{\label{sec:resultsaveraging}Inverse relations}

In Section \ref{sec:the Pseudopotential} we noticed the fact that
once either the Galilean or gauge invariance is imposed, the numbers
of parameters of the pseudopotential are the same, at each order, as
the numbers of coupling constants of the EDF for {\em each} isospin.
This situation allows us to obtain the inverse relations, namely,
expressions relating the coupling constants of the EDF to the
parameters of the pseudopotential. For the case of gauge invariance,
at second order they are given in Tables \ref{inverted second energy
isoscalar} and \ref{inverted second energy isovector}, at fourth
order in Tables \ref{inverted fourth energy isoscalar} and
\ref{inverted fourth energy isovector}, and at sixth order in Tables
\ref{inverted sixth energy isoscalar} and \ref{inverted sixth energy
isovector}. As sets of independent coupling constants of the
gauge-invariant EDF we selected the ones used in Appendix C of
Ref.~\cite{Carlsson1}. Note that in each case, the parameters of the
pseudopotential can be expressed either by the isoscalar or by the
isovector coupling constants. For the case of Galilean invariance,
analogous expressions are available in the supplemental material~\cite{suppl}.

\begin{table}
\caption{\label{inverted second energy isoscalar}Second-order
parameters of the pseudopotential as functions of the coupling
constants of the isoscalar EDF $(t=0)$ when the gauge invariance is imposed,
according to the formula
$C_{\tilde{n} \tilde{L},v_{12} S}^{\tilde{n}' \tilde{L}'}= A (
a C_{20,0000}^{0000,0}+
b C_{20,0011}^{0011,0}+
c C_{22,0011}^{0011,0}+
d C_{00,1101}^{1101,0}+
e C_{11,0011}^{1101,0}+
f C_{00,2011}^{0011,0}+
g C_{00,2211}^{0011,0})$.}
\begin{ruledtabular}
\begin{tabular}{ccccccccc}
& $A$  & $a$ & $b$ & $c$ & $d$ & $e$ & $f$ & $g$ \\
\hline
$C_{00,00}^{20}$       &    $-\frac{2}{3}$          &       10       &      2$\sqrt{3}$        &       0       &      5        &       0       & $-$$\sqrt{3}$      & 0  \\
$C_{00,20}^{20}$       &     $\frac{2}{3}$        &      $2\sqrt{3}$        &        6        &       0       &         $\sqrt{3}$     &       0       &  $-$3      &  0 \\
$C_{00,22}^{22}$       &     $-4$          &      0        &     0         &    $-$2         &        0       &   0      &       0    &   $1$  \\
$C_{11,00}^{11}$       &    $-\frac{2}{3}$           &   $-$6         &     $2\sqrt{3}$         &   0         &        3      &        0      &      $\sqrt{3}$     & 0  \\
$C_{11,20}^{11}$       &    $-\frac{2}{3}$             & $-2\sqrt{3}$       &    10       &      0     &    $\sqrt{3}$   &   0    &   5  &     0    \\
$C_{11,11}^{11}$       &     $-\frac{4}{3}$           &    0          &    0          &        0      &       0       &      1        &    0   &  0\\
$C_{11,22}^{11}$       &     $-\frac{4}{3}$           &      0        &     0         &        2      &         0     &        0      &    0   & 1 \\
\end{tabular}
\end{ruledtabular}
\end{table}

\begin{table}
\caption{\label{inverted second energy isovector}Same as in
Table~\ref{inverted second energy isoscalar} but for the isovector
EDF $(t=1)$, according to the formula
$C_{\tilde{n} \tilde{L},v_{12} S}^{\tilde{n}' \tilde{L}'}= A (
a C_{20,0000}^{0000,1}+
b C_{20,0011}^{0011,1}+
c C_{22,0011}^{0011,1}+
d C_{00,1101}^{1101,1}+
e C_{11,0011}^{1101,1}+
f C_{00,2011}^{0011,1}+
g C_{00,2211}^{0011,1})$.}
\begin{ruledtabular}
\begin{tabular}{ccccccccc}
   & $A$  & $a$ & $b$ & $c$ & $d$ & $e$ & $f$ & $g$ \\
\hline
$C_{00,00}^{20}$       &    $\frac{2}{3}$          &       $2\sqrt{3}$       &   6        &       0       &      $\sqrt{3}$        &       0       & $-$3      & 0  \\
$C_{00,20}^{20}$       &     $-\frac{2}{3}$        &      6         &        $-2\sqrt{3}$       &       0       &         3     &       0       &    $\sqrt{3}$      &  0 \\
$C_{00,22}^{22}$       &     $\frac{4}{3}$          &      0        &     0         &      $-2\sqrt{3}$         &        0       &   0      &       0    &   $\sqrt{3}$  \\
$C_{11,00}^{11}$       &    $-\frac{2}{3}$           &   $-$2$\sqrt{3}$         &   $-$6         &   0         &        $\sqrt{3}$     &        0      &    $-$3     & 0  \\
$C_{11,20}^{11}$       &    $\frac{2}{\sqrt{3}}$             &$-$2$\sqrt{3}$       &    2       &      0     &    $\sqrt{3}$   &   0    & 1  &     0    \\
$C_{11,11}^{11}$       &     $-\frac{4}{\sqrt{3}}$           &    0          &    0          &        0      &       0       &      1        &    0   &  0\\
$C_{11,22}^{11}$       &     $-\frac{4}{\sqrt{3}}$           &      0        &     0         &        2      &         0     &        0      &    0   & 1 \\
\end{tabular}
\end{ruledtabular}
\end{table}

\begin{table}
\caption{\label{inverted fourth energy isoscalar}Same as in
Table~\ref{inverted second energy isoscalar} but for the fourth-order
parameters of the pseudopotential, according to the formula
$C_{\tilde{n} \tilde{L},v_{12} S}^{\tilde{n}' \tilde{L}'}= A (
a C_{00,2202}^{2202,0}+
b C_{00,2212}^{2212,0}+
c C_{00,4211}^{0011,0}+
d C_{40,0000}^{0000,0}+
e C_{40,0011}^{0011,0}+
f C_{42,0011}^{0011,0})$.}
\begin{ruledtabular}
\begin{tabular}{cccccccc}
   & $A$  & $a$ & $b$ & $c$ & $d$ & $e$ & $f$  \\
\hline
$C_{11,00}^{31}$       & $\frac{2}{15}$ &  $18\sqrt{5}$   &    $-18\sqrt{5}$    &     $-7\sqrt{15}$       &      $-120$         &        $40\sqrt{3}$       &     $0$          \\
$C_{11,20}^{31}$       &  $\frac{2}{15}$ &  $6\sqrt{15}$   &    $-30\sqrt{15}$    &     $-35\sqrt{5}$       &      $-40\sqrt{3}$         &        $200$       &     $0$          \\
$C_{11,22}^{33}$       & $\frac{8}{3}\sqrt{\frac{7}{15}}$  & 0        &      0        &     $-1$          &      0        &        0       &     $2$     \\
$C_{22,00}^{22}$       & $\frac{1}{9}$ &  $30$             & 18       &     $7\sqrt{3} $        &      $40\sqrt{5}$     &    $8\sqrt{15}$   &  $0$         \\
$C_{22,20}^{22}$       & $-\frac{1}{9\sqrt{5}}$ &   $6\sqrt{15}$           &    $18\sqrt{15}$          &    $21\sqrt{5} $           &        $40\sqrt{3}$      &       120       &     $0$         \\
$C_{22,22}^{22}$   & $-\frac{4}{3}\sqrt{7}$   &  0   &    0  &     $1$       &    0        &       0       &     $2$        \\
\end{tabular}
\end{ruledtabular}
\end{table}

\begin{table}
\caption{\label{inverted fourth energy isovector}Same as in
Table~\ref{inverted second energy isovector} but for the fourth-order
parameters of the pseudopotential, according to
the formula
$C_{\tilde{n} \tilde{L},v_{12} S}^{\tilde{n}' \tilde{L}'}= A (
a C_{00,2202}^{2202,1}+
b C_{00,2212}^{2212,1}+
c C_{00,4211}^{0011,1}+
d C_{40,0000}^{0000,1}+
e C_{40,0011}^{0011,1}+
f C_{42,0011}^{0011,1})$.}
\begin{ruledtabular}
\begin{tabular}{cccccccc}
   & $A$  & $a$ & $b$ & $c$ & $d$ & $e$ & $f$  \\
\hline
$C_{11,00}^{31}$       &  $-\frac{2}{15}$  &  $-6\sqrt{15}$   &    $-18\sqrt{15}$    &     $-21\sqrt{5}$       &      $40\sqrt{3}$         &        $120$       &      $0$          \\
$C_{11,20}^{31}$       &  $-\frac{2}{15}$   &  $18\sqrt{5}$   &    $-18\sqrt{5}$    &     $-7\sqrt{15}$       &      $-120$         &     $40\sqrt{3}$           &      $0$          \\
$C_{11,22}^{33}$       &  $\frac{8}{3}\sqrt{\frac{7}{5}}$   & 0        &      0        &     $-1$          &      0        &        0       &     $2$     \\
$C_{22,00}^{22}$       &  $\frac{1}{9}$  & $-6\sqrt{3}$             & $-18\sqrt{3}$       &     $-21 $        &     $-8\sqrt{15}$      &   $-24\sqrt{5}$    &  $0$         \\
$C_{22,20}^{22}$       &    $\frac{1}{9\sqrt{5}}$  & $18\sqrt{5}$           &   $-18\sqrt{5}$           &    $-7\sqrt{15}$           &        120      &        $-40\sqrt{3}$        &     $0$         \\
$C_{22,22}^{22}$   &  $\frac{4}{3}\sqrt{\frac{7}{3}}$    & 0   &    0  &     $1$       &    0        &       0       &     $2$        \\
\end{tabular}
\end{ruledtabular}
\end{table}

\begin{table}
\caption{\label{inverted sixth energy isoscalar}Same as in
Table~\ref{inverted second energy isoscalar} but for the sixth-order
parameters of the pseudopotential, according to the formula
$C_{\tilde{n} \tilde{L},v_{12} S}^{\tilde{n}' \tilde{L}'}= A (
a C_{00,4212}^{2212,0}+
b C_{00,3303}^{3303,0}+
c C_{00,6211}^{0011,0}+
d C_{60,0000}^{0000,0}+
e C_{60,0011}^{0011,0}+
f C_{62,0011}^{0011,0})$.}
\begin{ruledtabular}
\setlength\tabcolsep{0.75pt}
\begin{tabular}{cccccccc}
   & $A$  & $a$ & $b$ & $c$ & $d$ & $e$ & $f$  \\
\hline
$C_{11,22}^{53}$ & $-\frac{16}{3}\sqrt{\frac{7}{15}}$ & 0 & 0 & $1$ & 0 & 0 & $2$\\
$C_{22,00}^{42}$       &  $\frac{2}{21}$   & $21$    &  $-15\sqrt{105}$     &     $42\sqrt{3}$       &     $-208\sqrt{5}$         &     $-56\sqrt{15}$          &     $0$  \\
$C_{22,20}^{42}$   &  $-\frac{2}{3}\sqrt{\frac{1}{7}}$   & $3{\sqrt{21}}$    &  $-9\sqrt{5}$     &     $18{\sqrt{7}}$       &     $-8{\sqrt{105}}$         &     $-24{\sqrt{35}}$          &     $0$ \\
$C_{22,22}^{44}$       &  $-\frac{16}{\sqrt{5}}$   & 0     &      0    &   $1$       &      0       &     0    &       $-2$   \\
$C_{33,00}^{33}$  &   $-\frac{2}{45}$   & $\sqrt{105}$    &  $45$     &     $6\sqrt{35}$       &     $-40\sqrt{21}$         &     $40\sqrt{7}$          &     $0$   \\
$C_{33,20}^{33}$  & $\frac{2}{9}\sqrt{\frac{1}{15}}$    & $-5\sqrt{21}$    &  $-9\sqrt{5}$     &     $-30\sqrt{7}$       &     $8\sqrt{105}$         &     $-40\sqrt{35}$ &     $0$  \\
\end{tabular}
\end{ruledtabular}
\end{table}

\begin{table}
\caption{\label{inverted sixth energy isovector}Same as in
Table~\ref{inverted second energy isovector} but for the sixth-order
parameters of the pseudopotential, according to
the formula
$C_{\tilde{n} \tilde{L},v_{12} S}^{\tilde{n}' \tilde{L}'}= A (
a C_{00,4212}^{2212,1}+
b C_{00,3303}^{3303,1}+
c C_{00,6211}^{0011,1}+
d C_{60,0000}^{0000,1}+
e C_{60,0011}^{0011,1}+
f C_{62,0011}^{0011,1})$.}
\begin{ruledtabular}
\setlength\tabcolsep{1pt}
\begin{tabular}{cccccccc}
   & $A$  & $a$ & $b$ & $c$ & $d$ & $e$ & $f$  \\
\hline
$C_{11,22}^{53}$       &   $-\frac{16}{3}\sqrt{\frac{7}{5}}$    &  0     &      0    &   $1$       &      0       &     0    &       $2$        \\
$C_{22,00}^{42}$       &   $\frac{2}{7\sqrt{3}}$   &  $-21$    &  $3\sqrt{105}$     &     $-42\sqrt{3}$       &     $56\sqrt{5}$         &     $56\sqrt{15}$          &     $0$          \\
$C_{22,20}^{42}$   &   $-\frac{2}{21}$   & $21$    &  $9\sqrt{105}$     &     $42\sqrt{3}$       &     $168\sqrt{5}$         &     $-56\sqrt{15}$          &     $0$         \\
$C_{22,22}^{44}$       &    $\frac{16}{\sqrt{15}}$     & 0     &      0    &   $1$       &      0       &     0    &       $-2$           \\
$C_{33,00}^{33}$  &   $-\frac{2}{45}$   & $-3\sqrt{35}$    &  $15\sqrt{3}$     &     $-6\sqrt{105}$       &     $-40\sqrt{7}$         &     $-40\sqrt{21}$          &     $0$             \\
$C_{33,20}^{33}$ &  $\frac{2}{9}\sqrt{\frac{1}{15}}$    &  $3\sqrt{7}$    &  $9\sqrt{15}$     &     $6\sqrt{21}$       &     $-24\sqrt{35}$         &     $8\sqrt{105}$          &     $0$         \\
\end{tabular}
\end{ruledtabular}
\end{table}

\subsection{\label{iso}Constraints on the Energy Density Functional}

The zero range of the pseudopotential is at the origin of the
specific constraints induced upon the resulting coupling constants of
the EDF. Indeed, elimination of the pseudopotential parameters from
pairs of relationships defined by Tables~\ref{inverted second energy
isoscalar}--\ref{inverted second energy isovector}, \ref{inverted
fourth energy isoscalar}--\ref{inverted fourth energy isovector}, and
\ref{inverted sixth energy isoscalar}--\ref{inverted sixth energy
isovector} leaves us with sets of linear equations that the EDF
coupling constants must obey. At second order, that is, for the
standard Skyrme interaction, this fact is well known and allows us to
express the time-odd coupling constants through the time-even ones,
see Ref.~\cite{Dobaczewski1} for the complete set of expressions. We
do not yet know if the analogous property may hold at higher orders,
because this fact crucially depend on the arbitrary choice of the
independent coupling constants that define the Galilean or gauge
symmetries.

In the present paper, we derive the set of constraints on the EDF
coupling constants that can be obtained by inverting the relations
for the isovector coupling constants, given in Tables~\ref{inverted
second energy isovector}, \ref{inverted fourth energy isovector}, and
\ref{inverted sixth energy isovector}. This allows us to express, at
each order, the isovector coupling constants through the isoscalar
ones. For the case of gauge invariance, at second, fourth, and sixth
order, such relations are listed in Tables~\ref{second energy vect
iso}, \ref{fourth-isotable}, and \ref{sixth-isotable}, respectively.
For the case of Galilean invariance, analogous expressions are
available in the supplemental material~\cite{suppl}.

\begin{table}
\caption{\label{second energy vect iso}Constraints on the EDF that is
derived by averaging the second-order gauge-invariant
pseudopotential, expressed by the formula $C_{mI,\tilde{n} L \nu
J}^{\tilde{n}' L' \nu' J', 1}=
a C_{00,1101}^{1101,0}+
b C_{00,2011}^{0011,0}+
c C_{00,2211}^{0011,0}+
d C_{11,0011}^{1101,0}+
e C_{20,0000}^{0000,0}+
f C_{20,0011}^{0011,0}+
g C_{22,0011}^{0011,0}$.}
\begin{ruledtabular}
\begin{tabular}{cccccccc}
   & $a$ & $b$ & $c$ & $d$ & $e$ & $f$ &$g$ \\
\hline
$C_{00,1101}^{1101,1}$      &    $-\frac{1}{\sqrt{3}}$          &       0       &      0        &       0     &   $-\frac{2}{\sqrt{3}}$            &  $-2$        & 0  \\
$C_{00,2011}^{0011,1}$       &     $0$        &     $-\frac{1}{\sqrt{3}}$          &       0       &       0       &     $2$          &       $-\frac{2}{\sqrt{3}}$        &      0   \\
$C_{00,2211}^{0011,1}$      &       0       &     0     &    $-\frac{1}{\sqrt{3}}$        &     0         &      0      &    0    &    $\frac{4}{\sqrt{3}}$          \\
$C_{11,0011}^{1101,1}$    &   0        &       0       &      0        &    $\frac{1}{\sqrt{3}}$            &       0       &     0        &      0   \\
$C_{20,0000}^{0000,1}$     &     $-\frac{1}{2\sqrt{3}}$           &  $\frac{1}{2}$    &    0      &      0    &    $-\frac{1}{\sqrt{3}}$      &   $0$     &   0  \\
$C_{20,0011}^{0011,1}$    &       $-\frac{1}{2}$         &     $-\frac{1}{2\sqrt{3}}$          &        0      &        0      &        0      &         $-\frac{1}{\sqrt{3}}$      &    0   \\
$C_{22,0011}^{0011,1}$     &    0          &    0          &   $\frac{1}{\sqrt{3}}$           &       0       &      0        &       0       &  $-\frac{1}{\sqrt{3}}$      \\
\end{tabular}
\end{ruledtabular}
\end{table}

\begin{table}
\caption{\label{fourth-isotable} Same as in Table~\ref{second energy
vect iso} but for the fourth-order terms, according to the formula
$C_{mI,n L v J}^{n' L' v' J', 1}= A(
a C_{00,2202}^{2202,0}+
b C_{00,2212}^{2212,0}+
c C_{00,4211}^{0011,0}+
d C_{40,0000}^{0000,0}+
e C_{40,0011}^{0011,0}+
f C_{42,0011}^{0011,0})$.}
\begin{ruledtabular}
\setlength\tabcolsep{0.75pt}
\begin{tabular}{cccccccc}
 &  $A$  & $a$ & $b$ & $c$ & $d$ & $e$ & $f$ \\
\hline
$C_{40,0000}^{0000,1}$  &  $\frac{1}{120}$  &    $-6\sqrt{15}$          &   $-18\sqrt{15}$             &      $-21\sqrt{5}$        &       $-40\sqrt{3}$     &   $0$            &  $0$         \\
$C_{40,0011}^{0011,1}$   &  $\frac{1}{120}$    &    $-18\sqrt{5}$         &   $18\sqrt{5}$          &      $7\sqrt{15}$          &       $0$      &     $-40\sqrt{3}$          &      $0$          \\
$C_{42,0011}^{0011,1}$   & $-\frac{1}{\sqrt{3}}$    &       0       &     0     &    $1$        &     0         &      0      &         $1$       \\
$C_{00,2202}^{2202,1}$   & $\frac{1}{9}$   &    $-3\sqrt{3}$        &    $0$           &      $0$        &       $-4\sqrt{15}$        &     $-12\sqrt{5}$         &      $0$           \\
$C_{00,4211}^{0011,1}$ &  $-\frac{1}{\sqrt{3}}$    &    0         &  0   &    $1$      &      0    &   0    &   $4$     \\
$C_{00,2212}^{2212,1}$    &  $\frac{1}{9}$    &       $0$         &      $-3\sqrt{3}$      &         $0$      &       $-4\sqrt{15}$         &      $4\sqrt{5}$         &       $14$   \\
\end{tabular}
\end{ruledtabular}
\end{table}

\begin{table}
\caption{\label{sixth-isotable}Same as in Table~\ref{second energy
vect iso} but for the sixth-order terms, according to the formula
$C_{mI,n L v J}^{n' L' v' J', 1}=A(
a C_{00,4212}^{2212,0}+
b C_{00,3303}^{3303,0}+
c C_{00,6211}^{0011,0}+
d C_{60,0000}^{0000,0}+
e C_{60,0011}^{0011,0}+
f C_{62,0011}^{0011,0})$.}
\begin{ruledtabular}
\setlength\tabcolsep{0.4pt}
\begin{tabular}{cccccccc}
 &   $A$     & $a$ & $b$ & $c$ & $d$ & $e$ & $f$ \\
\hline
$C_{60,0000}^{0000,1}$      &   $\frac{1}{840}$  &     $21\sqrt{15}$          &   $-45\sqrt{7}$             &  $126\sqrt{5}$         &        $-280\sqrt{3}$     &   $0$            &  $0$         \\
$C_{60,0011}^{0011,1}$       &   $\frac{1}{840}$   &  $-21\sqrt{5}$         &   $-45\sqrt{21}$          &      $-42\sqrt{15}$          &       0      &     $-280\sqrt{3}$          &      $0$          \\
$C_{62,0011}^{0011,1}$      &   $\frac{1}{\sqrt{3}}$    &   0       &     0     &    $1$        &     0         &      0      &         $-1$       \\
$C_{00,3303}^{3303,1}$    &  $\frac{1}{9}$       & $0$        &    $-3\sqrt{3}$           &      $0$        &   $-8\sqrt{7}$           &       $-8\sqrt{21}$       &      $0$           \\
$C_{00,6211}^{0011,1}$     &   $-\frac{1}{\sqrt{3}}$     &   0         &  0   &    $1$      &      0    &   0    &   $-4$     \\
$C_{00,4212}^{2212,1}$    &     $\frac{1}{3}$    &    $-\sqrt{3}$         &    $0$           &         $0$       &        $8\sqrt{15}$      &    $-8\sqrt{5}$          &       $-24$           \\
\end{tabular}
\end{ruledtabular}
\end{table}

\section{\label{spherical}Relations between the pseudopotential and
Energy Density Functional with conserved spherical symmetry}

In this Section, we assume the spherical, space-inversion, and
time-reversal symmetries of the EDF, see Sec.~IV of
Ref.~\cite{Carlsson1}. In this way we make our results applicable to
the simplest case of spherical even-even nuclei. Below we fully show
explicit results for the case of gauge symmetry conserved,
whereas the full results pertaining to the case of Galilean symmetry
are given in the supplemental material~\cite{suppl}.

When the gauge symmetry is imposed on the EDF and the isospin degree
of freedom is taken into account, we have 8 independent spherical EDF
terms at second order, 6 at fourth order, and 6 at sixth order. The 8
corresponding second-order coupling constants can then be expressed
by the 7 second-order pseudopotential parameters. Similarly, both at
fourth and sixth orders, 6 coupling constants can then be expressed by
6 pseudopotential parameters.

As is well known, at second order the isoscalar and isovector spin-orbit coupling
constants depend both on one spin-orbit pseudopotential parameter,
namely,
\begin{subequations}
\begin{eqnarray}
C_{11,1111}^{0000,0} &=& -\frac{3}{4}C_{11,11}^{11}   , \\
C_{11,1111}^{0000,1} &=& -\frac{\sqrt{3}}{4}C_{11,11}^{11} ,
\end{eqnarray}
\end{subequations}
which gives one constraint on the spin-orbit coupling constants,
\begin{equation}
C_{11,1111}^{0000,1}=\frac{1}{\sqrt{3}}C_{11,1111}^{0000,0} .
\label{eq:0sph}
\end{equation}
The remaining 6 spherical EDF coupling constants expressed
through 6 pseudopotential parameters are given
in Table~\ref{second energy spherical}. Similar expressions
at fourth and sixth orders are given in
Tables~\ref{fourth energy spherical} and~\ref{sixth energy
spherical}. As in Sec.~\ref{sec:resultsaveraging}, from these results
we can obtain the inverse expressions relating the parameters of the
pseudopotential to the coupling constants of the spherical EDF; these
are given in Tables~\ref{inverted second energy
spherical}--\ref{inverted sixth energy spherical}.

\begin{table}
\caption{\label{second energy spherical}Second-order coupling
constants of the EDF as functions of parameters of the
pseudopotential when the gauge and the spherical symmetries are simultaneously imposed, according to the formula $C_{mI,n L v J}^{n' L' v'
J', t}= A(a C_{00,00}^{20}+b C_{00,20}^{20}+c C_{00,22}^{22}+d
C_{11,00}^{11}+e C_{11,20}^{11}+f C_{11,22}^{11})$.}
\begin{ruledtabular}
\begin{tabular}{cccccccc}
& $A$ & $a$ & $b$ & $c$ & $d$ & $e$ & $f$ \\
\hline
$C_{20,0000}^{0000,0}$       &    $\frac{1}{32}$          &     $-$3       &      $-\sqrt{3}$ &       0       &   5        &  $-\sqrt{3}$     & 0  \\
$C_{20,0000}^{0000,1}$       &    $\frac{1}{32}$          &   $\sqrt{3}$   &      $-3$        &       0       & $\sqrt{3}$ &     $-3$         & 0  \\
$C_{00,2000}^{0000,0}$       &    $\frac{1}{16}$          &        3       &      $\sqrt{3}$  &       0       &   5        &  $-\sqrt{3}$     &  0 \\
$C_{00,2000}^{0000,1}$       &    $\frac{1}{16}$          &   $-\sqrt{3}$  &      $3$         &       0       & $\sqrt{3}$ &     $-3$         &  0 \\
$C_{00,1111}^{1111,0}$       &    $\frac{1}{48}$          &        3       &      $5\sqrt{3}$ &  $-\sqrt{15}$ &$-$3        &  $3\sqrt{3}$     & $-3\sqrt{15}$   \\
$C_{00,1111}^{1111,1}$       &    $\frac{1}{16}$          &   $\sqrt{3}$   &       1          &  $\sqrt{5}$   & $-\sqrt{3}$&    $-$1          & $-\sqrt{5}$   \\
\end{tabular}
\end{ruledtabular}
\end{table}

\begin{table}
\caption{\label{fourth energy spherical}Same as in
Table~\ref{second energy spherical} but for the fourth-order coupling
constants of the EDF, according to the formula $C_{mI,n L v J}^{n' L' v'
J', t}= A(a C_{11,00}^{31}+b C_{11,20}^{31}+c C_{11,22}^{33}+d
C_{22,00}^{22}+e C_{22,20}^{22}+f C_{22,22}^{22})$.}
\begin{ruledtabular}
\begin{tabular}{cccccccc}
& $A$ & $a$ & $b$ & $c$ & $d$ & $e$ & $f$  \\
\hline
$C_{40,0000}^{0000,0}$       &    $\frac{1}{640}$          &     $-$25       &      $5\sqrt{3}$        &       0       &    $18\sqrt{5}$           &     $6\sqrt{15}$          &   0     \\
$C_{40,0000}^{0000,1}$       &    $\frac{1}{640}$          &      $-5\sqrt{3}$        &      15        &       0       &       $-6\sqrt{15}$       &     $18\sqrt{5}$          &   0     \\
$C_{00,2202}^{2202,0}$       &     $\frac{1}{96}$        &      $5\sqrt{5}$          &      $-\sqrt{15}$         &       0       &         18     &      $6\sqrt{3}$         &     0     \\
$C_{00,2202}^{2202,1}$       &     $\frac{1}{96}$        &     $\sqrt{15}$         &   $-3\sqrt{5}$           &       0       &     $-6\sqrt{3}$          &     18        &     0    \\
$C_{00,3111}^{1111,0}$       &    $\frac{1}{80}$           &   $-$5         &     $5\sqrt{3}$          &      $-15\sqrt{7}$          &     $6\sqrt{5}$         &   $10\sqrt{15}$           &     $-2\sqrt{105}$      \\
$C_{00,3111}^{1111,1}$       &    $\frac{1}{80}$           &     $-5\sqrt{3}$         &   $-$5          &      $-5\sqrt{21}$          &      $6\sqrt{15}$        &      $6\sqrt{5}$         &  $6\sqrt{35}$   \\
\end{tabular}
\end{ruledtabular}
\end{table}

\begin{table}
\caption{\label{sixth energy spherical}Same as in
Table~\ref{second energy spherical} but for the sixth-order coupling
constants of the EDF, according to the formula $C_{mI,n L v J}^{n' L' v'
J', t}= A(a C_{11,22}^{53}+b C_{22,00}^{42}+c C_{22,20}^{42}+d
C_{22,22}^{44}+e C_{33,00}^{33}+f C_{33,20}^{33})$.}
\begin{ruledtabular}
\setlength\tabcolsep{0.2pt}
\begin{tabular}{cccccccc}
& $A$ & $a$ & $b$ & $c$ & $d$ & $e$ & $f$  \\
\hline
$C_{60,0000}^{0000,0}$       &    $\frac{1}{4480}$          &      0      &   $-21\sqrt{5}$   &   $-7\sqrt{15}$      &  0        &     $75\sqrt{21}$          &   $-45\sqrt{7}$     \\
$C_{60,0000}^{0000,1}$       &    $\frac{1}{4480}$          &    0     &    $7\sqrt{15}$         &       $-21\sqrt{5}$       &    0      &     $45\sqrt{7}$          &  $-45\sqrt{21}$       \\
$C_{00,6000}^{0000,0}$       &     $\frac{1}{2240}$        &      0         &      $21\sqrt{5}$         &      $7\sqrt{15}$        &       0    &      $75\sqrt{21}$      &   $-45\sqrt{7}$         \\
$C_{00,6000}^{0000,1}$       &     $\frac{1}{2240}$        &     0        &   $-7\sqrt{15}$           &      $21\sqrt{5}$           &     0          &    $45\sqrt{7}$           &    $-45\sqrt{21}$      \\
$C_{00,3111}^{3111,0}$       &    $\frac{1}{800}$           &    $-135\sqrt{7}$            &     $21\sqrt{5}$          &      $35\sqrt{15}$          &     $-105\sqrt{3}$         &   $-45\sqrt{21}$           &     $135\sqrt{7}$      \\
$C_{00,3111}^{3111,1}$       &    $-\frac{3}{800}$           &     $15\sqrt{21}$         &  $-7\sqrt{15}$                 &      $-7\sqrt{5}$          &    $-$105       &      $45\sqrt{7}$         &  $15\sqrt{21}$   \\
\end{tabular}
\end{ruledtabular}
\end{table}

\begin{table}
\caption{\label{inverted second energy spherical}Second-order
parameters of the pseudopotential (spin-orbit term not included) as functions of the coupling
constants of the EDF when the gauge and the spherical symmetries are simultaneously imposed,
according to the formula
$C_{\tilde{n} \tilde{L},v_{12} S}^{\tilde{n}' \tilde{L}'}=
a C_{20,0000}^{0000,0}+
b C_{20,0000}^{0000,1}+
c C_{00,2000}^{0000,0}+
d C_{00,2000}^{0000,1}+
e C_{00,1111}^{1111,0}+
f C_{00,1111}^{1111,1}$.}
\begin{ruledtabular}
\begin{tabular}{ccccccc}
& $a$  & $b$ & $c$ & $d$ & $e$ & $f$ \\
\hline
$C_{00,00}^{20}$       &  $-$4          &    $\frac{4}{\sqrt{3}}$          &      2       &  $-\frac{2}{\sqrt{3}}$            &      0        &       0       \\
$C_{00,20}^{20}$       &      $-\frac{4}{\sqrt{3}}$      &   $-$4        &    $\frac{2}{\sqrt{3}}$            &       2       &        0     &       0      \\
$C_{00,22}^{22}$       &      $\frac{16}{\sqrt{15}}$        &     $-\frac{16}{\sqrt{5}}$         &       $\frac{4}{\sqrt{15}}$        &      $-\frac{4}{\sqrt{5}}$          &     $-4\sqrt{\frac{3}{5}}$          &  $\frac{12}{\sqrt{5}}$      \\
$C_{11,00}^{11}$       &   4          &     $-\frac{4}{\sqrt{3}}$         &     2        &   $-\frac{2}{\sqrt{3}}$          &        0      &        0       \\
$C_{11,20}^{11}$       &     $\frac{4}{\sqrt{3}}$          &   $-\frac{20}{3}$     &    $\frac{2}{\sqrt{3}}$       &    $-\frac{10}{3}$       &    0   &   0      \\
$C_{11,22}^{11}$       &    $-\frac{16}{\sqrt{15}}$            &     $-\frac{16}{3\sqrt{5}}$           &     $\frac{4}{\sqrt{15}}$           &   $\frac{4}{3\sqrt{5}}$              &   $-4\sqrt{\frac{3}{5}}$   &   $-\frac{4}{\sqrt{5}}$   \\
\end{tabular}
\end{ruledtabular}
\end{table}

\begin{table}
\caption{\label{inverted fourth energy spherical}Fourth-order
parameters of the pseudopotential as functions of the coupling
constants of the EDF when the gauge and the spherical symmetries are simultaneously imposed, according to
the formula
$C_{\tilde{n} \tilde{L},v_{12} S}^{\tilde{n}' \tilde{L}'}=
a C_{40,0000}^{0000,0}+
b C_{40,0000}^{0000,1}+
c C_{00,2202}^{2202,0}+
d C_{00,2202}^{2202,1}+
e C_{00,3111}^{1111,0}+
f C_{00,3111}^{1111,1}$.}
\begin{ruledtabular}
\begin{tabular}{ccccccc}
   & $a$ & $b$ & $c$ & $d$ & $e$ & $f$  \\
\hline
$C_{11,00}^{31}$       &  $-16$  &  $\frac{16}{\sqrt{3}}$  &   $\frac{12}{\sqrt{5}}$     &     $-4\sqrt{\frac{3}{5}}$       &      0      &     0              \\
$C_{11,20}^{31}$       & $-\frac{16}{\sqrt{3}}$  & $\frac{80}{3}$  &  $4\sqrt{\frac{3}{5}}$    &     $-4\sqrt{5}$       &      0        &     0       \\
$C_{11,22}^{33}$       & $\frac{64}{3\sqrt{7}}$    &    $\frac{64}{3\sqrt{21}}$      &    $\frac{8}{\sqrt{35}}$           &    $\frac{8}{\sqrt{105}}$         &     $-\frac{4}{\sqrt{7}}$         &  $-\frac{4}{\sqrt{21}}$           \\
$C_{22,00}^{22}$       &  $\frac{8\sqrt{5}}{3}$  &    $-\frac{8}{3}\sqrt{\frac{5}{3}}$          & 2      &     $-\frac{2}{\sqrt{3}}$     &     0     &   0        \\
$C_{22,20}^{22}$       &  $\frac{8}{3}\sqrt{\frac{5}{3}}$    &    $\frac{8\sqrt{5}}{3}$        &   $\frac{2}{\sqrt{3}}$            &   2         &       0     &  0       \\
$C_{22,22}^{22}$   &  $-\frac{32}{3}\sqrt{\frac{5}{21}}$    &  $\frac{32}{3}\sqrt{\frac{5}{7}}$   &    $\frac{4}{\sqrt{21}}$   &   $-\frac{4}{\sqrt{7}}$       &     $-2\sqrt{\frac{5}{21}}$        &   $2\sqrt{\frac{5}{7}}$                 \\
\end{tabular}
\end{ruledtabular}
\end{table}

\begin{table}
\caption{\label{inverted sixth energy spherical}Same as in
Table~\ref{inverted fourth energy spherical} but for the sixth-order parameters of the pseudopotential, according to
the formula
$C_{\tilde{n} \tilde{L},v_{12} S}^{\tilde{n}' \tilde{L}'}=
a C_{60,0000}^{0000,0}+
b C_{60,0000}^{0000,1}+
c C_{00,6000}^{0000,0}+
d C_{00,6000}^{0000,1}+
e C_{00,3111}^{3111,0}+
f C_{00,3111}^{3111,1}$.}
\begin{ruledtabular}
\begin{tabular}{ccccccc}
   & $a$ & $b$ & $c$ & $d$ & $e$ & $f$  \\
\hline
$C_{11,22}^{53}$       & $-\frac{64\sqrt{7}}{9}$   &  $-\frac{64}{9}\sqrt{\frac{7}{3}}$  &   $\frac{16\sqrt{7}}{9}$     &    $\frac{16}{9}\sqrt{\frac{7}{3}}$       &    $-\frac{40}{9\sqrt{7}}$         &    $-\frac{40}{9\sqrt{21}}$                \\
$C_{22,00}^{42}$       & $-16\sqrt{5}$  & $16\sqrt{\frac{5}{3}}$  &  $8\sqrt{5}$  &   $-8\sqrt{\frac{5}{3}}$        &      0        &     0       \\
$C_{22,20}^{42}$       &  $-16\sqrt{\frac{5}{3}}$   &  $-16\sqrt{5}$     &    $8\sqrt{\frac{5}{3}}$          &     $8\sqrt{5}$        &    0        & 0       \\
$C_{22,22}^{44}$       &  $\frac{64}{3\sqrt{3}}$  &    $-\frac{64}{3}$          &  $\frac{16}{3\sqrt{3}}$     &    $-\frac{16}{3}$      &   $-\frac{40}{21\sqrt{3}}$        &   $\frac{40}{21}$         \\
$C_{33,00}^{33}$       &  $\frac{16}{3}\sqrt{\frac{7}{3}}$    &   $-\frac{16\sqrt{7}}{9}$           &    $\frac{8}{3}\sqrt{\frac{7}{3}}$         &     $-\frac{8}{9}\sqrt{7}$         &       0     &  0       \\
$C_{33,20}^{33}$   &  $\frac{16\sqrt{7}}{9}$   & $-\frac{80}{9}\sqrt{\frac{7}{3}}$    &  $\frac{8\sqrt{7}}{9}$      &   $-\frac{40}{9}\sqrt{\frac{7}{3}}$      &     0       &   0               \\
\end{tabular}
\end{ruledtabular}
\end{table}

At second order, the gauge and Galilean symmetries are
equivalent to one another~\cite{Carlsson1}. When at higher orders the Galilean invariance is imposed
on the spherical EDF, we have at fourth (sixth)
order 18 (32) independent terms, of which 4 (8) are of the spin-orbit
character. It turns out that, in the same way as for the second
order, the higher-order spin-orbit coupling constants are related
only to the spin-orbit pseudopotential parameters. Namely, at fourth
order we have
\begin{subequations}
\begin{eqnarray}
C_{31,1111}^{0000,0}&=&\frac{3}{16}C_{11,11}^{31}-\frac{1}{8}\sqrt{\frac{3}{5}}C_{22,11}^{22} , \\
C_{31,1111}^{0000,1}&=&\frac{1}{16}\sqrt{3}C_{11,11}^{31}+\frac{3}{8}\sqrt{\frac{1}{5}} C_{22,11}^{22} , \\
C_{11,3111}^{0000,0}&=&-\frac{3}{16}C_{11,11}^{31}-\frac{1}{8}\sqrt{\frac{3}{5}}C_{22,11}^{22}  , \\
C_{11,3111}^{0000,1}&=&-\frac{1}{16}\sqrt{3}C_{11,11}^{31}+\frac{3}{8}\sqrt{\frac{1}{5}} C_{22,11}^{22} ,
\end{eqnarray}
\end{subequations}
which gives the following constraints on the spin-orbit coupling constants:
\begin{subequations}
\label{eq:1sph}
\begin{eqnarray}
C_{31,1111}^{0000,1}&=&-\frac{1}{\sqrt{3}}C_{31,1111}^{0000,0} - \frac{2}{\sqrt{3}}C_{11,3111}^{0000,0}  ,  \label{subeq: 1of1sph}  \\
C_{11,3111}^{0000,1}&=&-\frac{2}{\sqrt{3}}C_{31,1111}^{0000,0}-\frac{1}{\sqrt{3}}C_{11,3111}^{0000,0} , \label{subeq: 2of1sph}
\end{eqnarray}
\end{subequations}
and at sixth order we have
\begin{subequations}
\begin{eqnarray}
\nonumber C_{51,1111}^{0000,0}=&&-\frac{3}{64}C_{11,11}^{51}+\frac{1}{32}\sqrt{\frac{3}{5}} C_{22,11}^{42}-\frac{3}{64}C_{31,11}^{31} \\
&&-\frac{27}{80}\sqrt{\frac{1}{14}}C_{33,11}^{33}, \\
\nonumber C_{51,1111}^{0000,1}=&&-\frac{\sqrt{3}}{64}C_{11,11}^{51}- \frac{3}{32}\sqrt{\frac{1}{5}}C_{22,11}^{42}-\frac{\sqrt{3}}{64}C_{31,11}^{31} \\
&&-\frac{9}{80}\sqrt{\frac{3}{14}}C_{33,11}^{33}, \\
\nonumber C_{11,5111}^{0000,0}=&& -\frac{3}{64}C_{11,11}^{51}-\frac{1}{32}\sqrt{\frac{3}{5}}C_{22,11}^{42} -\frac{3}{64}C_{31,11}^{31} \\
&&-\frac{27}{80}\sqrt{\frac{1}{14}}C_{33,11}^{33}, \\
\nonumber C_{11,5111}^{0000,1}=&&-\frac{\sqrt{3}}{64}C_{11,11}^{51}+\frac{3}{32}\sqrt{\frac{1}{5}}C_{22,11}^{42}-\frac{\sqrt{3}}{64}C_{31,11}^{31} \\
&&-\frac{9}{80}\sqrt{\frac{3}{14}}C_{33,11}^{33}, \\
\nonumber C_{31,3111}^{0000,0}=&&\frac{21}{160}C_{11,11}^{51}+\frac{9}{160}C_{31,11}^{31} \\ &&-\frac{27}{200}\sqrt{\frac{7}{2}}C_{33,11}^{33}, \\
\nonumber C_{31,3111}^{0000,1}=&&\frac{7}{160}\sqrt{3}C_{11,11}^{51} +\frac{3}{160}\sqrt{3}C_{31,11}^{31} \\
&& -\frac{9}{200}\sqrt{\frac{21}{2}}C_{33,11}^{33}, \\
\nonumber C_{33,3313}^{0000,0}=&&\frac{1}{24}\sqrt{\frac{7}{2}}C_{11,11}^{51}-\frac{1}{24}\sqrt{\frac{7}{2}}C_{31,11}^{31} \\
&& +\frac{3}{80}C_{33,11}^{33}, \\
\nonumber C_{33,3313}^{0000,1}=&&\frac{1}{24}\sqrt{\frac{7}{6}}C_{11,11}^{51}-\frac{1}{24}\sqrt{\frac{7}{6}}C_{31,11}^{31} \\
&& +\frac{\sqrt{3}}{80}C_{33,11}^{33} ,
\end{eqnarray}
\end{subequations}
which gives the constraints:
\begin{subequations}
\label{eq:2sph}
\begin{eqnarray}
C_{51,1111}^{0000,1}&=&-\frac{1}{\sqrt{3}}C_{51,1111}^{0000,0} + \frac{2}{\sqrt{3}}C_{11,5111}^{0000,0}  ,  \label{subeq: 1of2sph}  \\
C_{11,5111}^{0000,1}&=&\frac{2}{\sqrt{3}}C_{51,1111}^{0000,0}-\frac{1}{\sqrt{3}}C_{11,5111}^{0000,0} , \label{subeq: 2of2sph} \\
C_{31,3111}^{0000,1}&=&\frac{1}{\sqrt{3}}C_{31,3111}^{0000,0} ,  \label{subeq: 3of2sph}  \\
C_{33,3313}^{0000,1}&=&\frac{1}{\sqrt{3}}C_{33,3313}^{0000,0} .  \label{subeq: 4of2sph}
\end{eqnarray}
\end{subequations}

\begin{figure}
\includegraphics[width=7cm]{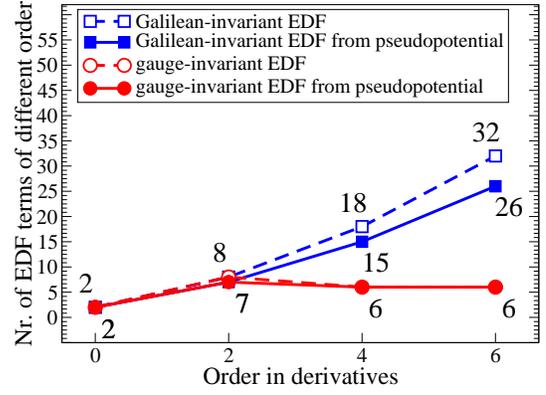}
\caption{\label{numtermEDF}(Color online) Number of terms of the spherical
EDF that is related to a pseudopotential (solid lines). Full squares and circles
show results for the Galilean and gauge invariance, respectively.
For reference, dashed lines with open squares and circles show the
corresponding results for the general spherical EDF studied in Ref.~\cite{Carlsson1}.}
\end{figure}

If now we consider the Galilean-invariant and spherical EDF without
spin-orbit terms, we obtain at fourth (sixth) order 1 (2) possible
constraints among the remaining 14 (24) coupling constants related to
the remaining 13 (22) parameters of the pseudopotential. These
results are available in the supplemental material~\cite{suppl}. Of
course, such constraints can be imposed in very many different ways.
We have checked that, in fact, not any of the 1 (2) coupling
constants of the fourth (sixth) order spherical EDF can be considered
as being dependent on all the other coupling constants. In the
supplemental material we present one example of a possible choice,
whereby at fourth (sixth) order the coupling constants $C_{22,1111}^{1111,1}$ ($C_{00,3111}^{3111,0}$
and $C_{00,3111}^{3111,1}$) are selected to be dependent.
A comparison between the numbers of terms of the Galilean-invariant and  gauge-invariant spherical
EDF with and without constraints coming from the reference to the pseudopotential is plotted in Fig.~\ref{numtermEDF}.

\section{\label{conclusions}Conclusions}

In summary, in this work we derived the Galilean-invariant nuclear
N$^{3}$LO pseudopotential with derivatives up to sixth order and
found the corresponding N$^{3}$LO EDF, which was obtained by
calculating the corresponding HF average energy. Owing to the
zero range of the pseudopotential, the number of terms thereof is
twice smaller then that of the most general EDF. We found explicit
linear relations between the parameters of the pseudopotential and
coupling constant of the EDF. These linear relations constitute a set
of constraints, which allow for expressing one half of the coupling
constants through the other half. As an example of such constraints,
we have derived linear relations between the isoscalar and isovector
coupling constants. The gauge-invariant form of the pseudopotential
was also derived, and all derivations were repeated also for this case.

\begin{table*}
\caption{\label{generalTable}Number of terms of different orders in
the pseudopotential (\ref{eq:1}) and in the EDF up to N$^{3}$LO,
evaluated for the conserved Galilean and gauge symmetries. The
last four columns show the number of terms in the EDF evaluated by taking
into account the additional constraints coming from the relation of
the EDF to pseudopotential.}
\begin{ruledtabular}
\begin{tabular}{ccccccccccc}
 &\multicolumn{2}{c}{Pseudopotential}&\multicolumn{8}{c}{EDF}\\
&&&\multicolumn{4}{c}{Not related to pseudopotential}&\multicolumn{4}{c}{Related to pseudopotential}\\
&&&\multicolumn{2}{c}{General}&\multicolumn{2}{c}{Spherical}&\multicolumn{2}{c}{General}&\multicolumn{2}{c}{Spherical}\\
 Order&Galilean&Gauge&Galilean&Gauge&Galilean&Gauge&Galilean&Gauge&Galilean&Gauge\\ \hline
 0&2&2&4&4&2&2&2&2&2&2\\
2&7&7&14&14&8&8&7&7&7&7  \\
4&15&6&30&12&18&6&15&6&15&6  \\
6&26&6&52&12&32&6&26&6&26&6  \\
\hline
N$^{3}$LO&50&21&100&42&60&22&50&21&50&21  \\
\end{tabular}
\end{ruledtabular}
\end{table*}

We have also analyzed properties of the EDF restricted by imposing
the spherical, space-inversion, and time-reversal symmetries, which
are relevant for describing spherical nuclei. In this case, by
relating the EDF to the pseudopotential, at second, fourth, and sixth
order one reduces the numbers of coupling constants only from 8, 18,
and 32 to 7, 15, and 26, respectively. Such reduction has two
origins: (i) at each order 1, 2, or 4 spin-orbit isovector and
isoscalar coupling constants become dependent on one another and (ii)
at fourth and sixth order one or two non-spin-orbit coupling
constants become linearly dependent on the remaining 13 or 22 ones,
respectively. Therefore, in spherical magic nuclei one can expect
relatively small effects related to imposing on the EDF the
pseudopotential origins, whereas this may have much more important
consequences in deformed, asymmetric, odd, and/or rotating nuclei.
We also note that for the EDF related to pseudopotential, imposing
the spherical symmetry does not change the numbers of idependent
coupling constants as compared to the general case.

Table~\ref{generalTable} gives an overview of the results by showing
the number of terms of pseudopotential and EDF with Galilean or
gauge symmetries imposed.

\acknowledgments

This work was supported in part by the Academy of Finland and the
University of Jyv\"askyl\"a within the FIDIPRO programme, and by the
Polish Ministry of Science and Higher Education under Contract
No.~N~N202~328234.

\appendix

\section{\label{appA}Time-reversal invariance and hermiticity of the pseudopotential}

The pseudopotential studied in this work is a contact interaction
built with derivative and spin operators. Furthermore, the choice
concerning the formalism is the use of the spherical tensors. Under
these assumptions, the general structure of the pseudopotential is
based on the following building blocks,
\begin{equation}
\hat{V}_0 =
\left[ \left[K'_{\tilde{n}'\tilde{L}'} K_{\tilde{n}\tilde{L}}\right]_{S}
\SSS_{v_{12} S}\right]_{0} \hat{\delta}_{12}(\bm{r}'_1\bm{r}'_2;\bm{r}_1\bm{r}_2)
\label{eq:0appA}.
\end{equation}

The final coupling to a scalar ensures that $\hat{V}_0$ is invariant
under space rotation Moreover, provided that $\tilde{n}'+\tilde{n}$
is even, it is also invariant under space-inversion. Now we proceed
to explore another fundamental symmetry, the time-reversal, and later
we also require the hermiticity of the pseudopotential.

The time-reversal operator $\hat{T}=-i\sigma_y\hat{K}$, where
$\hat{K}$ is the complex conjugation in space representation, can be
explicitly applied to the spherical-tensor representations of
momentum and spin operators, Eqs.~(\ref{eq:05a}), (\ref{eq:03a}), and
(\ref{eq:08a}), which gives the generic result for spherical tensors,
\begin{equation}
\label{eq:Trule}
\hat{T}A_{\lambda\mu}\hat{T}^\dagger = T_A(-1)^{\lambda-\mu}A_{\lambda,-\mu},
\end{equation}
where $T_A$ are numerical phase factors. In our case, we obtain $T_k=-1$
for the momentum operator and $T_{\sigma_v}=(-1)^v$ for the scalar ($v=0$)
and vector ($v=1$) spin operators. Moreover, since the Clebsh-Gordan coefficients
are real, rule (\ref{eq:Trule}) propagates through the angular
momentum coupling, that is, if phase factors $T_A$ and $T_{A'}$
characterize tensors $A_\lambda$ and  $A'_{\lambda'}$, respectively,
then the coupled tensor,
\begin{equation}
A''_{\lambda''\mu''} = [A_\lambda  A'_{\lambda'}]_{\lambda''\mu''} =
       \sum_{\mu\mu'} C^{\lambda''\mu''}_{\lambda\mu\lambda'\mu'}
       A_{\lambda\mu}A'_{\lambda'\mu'},
\end{equation}
is characterized by the product of phase factors $T_{A''}= T_A T_{A'}$
(cf.\ Appendix B in Ref.~\cite{Carlsson1}).
Therefore, the coupled operators appearing in $\hat{V}_0$ (\ref{eq:0appA}) are
characterized by the following values of phase factors,
\begin{subequations}
\label{eq:2appA}
\begin{eqnarray}
T_{K'_{\tilde{n}'\tilde{L}'}}&=&(-1)^{\tilde{n}'} , \label{subeq:1of2appA} \\
T_{K_ {\tilde{n }\tilde{L} }}&=&(-1)^{\tilde{n} } , \label{subeq:2of2appA} \\
T_{S_ {v_{12}S}}             &=&(-1)^{v_{12} } .    \label{subeq:3of2appA}
\end{eqnarray}
\end{subequations}
Finally, because the Dirac delta is real, for $\hat{V}_0$ we have,
\begin{equation}
T_{\hat{V}_0}=(-1)^{\tilde{n}'+\tilde{n }+v_{12}} ,
\label{eq:3appA}
\end{equation}
and by taking into account the space-inversion invariance, it boils down to
\begin{equation}
T_{\hat{V}_0}=(-1)^{v_{12}} .
\label{eq:4appA}
\end{equation}
This justifies the phase factor $i^{v_{12}}$ in the
definition of the pseudopotential in Eq.~(\ref{eq:1}), which ensures that
for real parameters, all terms of the pseudopotential are time-even.

Now we can proceed to calculate the adjoint
of the operator
$\hat{V}_0$ (\ref{eq:0appA}) multiplied by the phase factor derived
above, that is,
\begin{equation}
\left(i^{v_{12}}\hat{V}_0\right)^\dagger = (-i)^{v_{12}}
\left[ \left[K_{\tilde{n}'\tilde{L}'} K'_{\tilde{n}\tilde{L}}\right]_{S}^*
\SSS_{v_{12} S}^\dagger\right]_{0} \hat{\delta}_{12}(\bm{r}'_1\bm{r}'_2;\bm{r}_1\bm{r}_2),
\label{eq:9appA}
\end{equation}
where we treat the space derivatives of the Dirac delta like ordinary
numbers and the space variables had to be exchanged,
$\bm{r}'_1\leftrightarrow\bm{r}_1$ and
$\bm{r}'_2\leftrightarrow\bm{r}_2$.

Properties of generic spherical tensors under the complex
and Hermitian conjugations are given by the following rules,
\begin{equation}
\label{eq:Crule}
A_{\lambda\mu}^* = P_A(-1)^{\lambda-\mu}A_{\lambda,-\mu},
\end{equation}
\begin{equation}
\label{eq:Hrule}
A_{\lambda\mu}^\dagger = H_A(-1)^{\lambda-\mu}A_{\lambda,-\mu},
\end{equation}
where the phase factors $P_A$ and $H_A$ can be directly derived from
definitions~(\ref{eq:05a}), (\ref{eq:03a}), and (\ref{eq:08a}), that
is, $P_k=-1$ and $H_{\sigma_v}=+1$. These rules also propagate
through the angular
momentum coupling, that is, $P_{A''}= P_A P_{A'}$ and, for {\em commuting
operators}, which is the case here, $H_{A''}= H_A H_{A'}$.
Therefore, we have,
\begin{equation}
P_{\left[K_{\tilde{n}'\tilde{L}'} K'_{\tilde{n}\tilde{L}}\right]_{S}}
=(-1)^{\tilde{n}'+\tilde{n}}=+1 ,
\label{eq:13appA}
\end{equation}
and
\begin{equation}
H_{\SSS_{v_{12} S}}=+1 .
\label{eq:14appA}
\end{equation}
Finally, the adjoint operator of Eq.~(\ref{eq:9appA}) is given by
\begin{widetext}
\begin{equation}
\left(i^{v_{12}}\hat{V}_0\right)^\dagger = (-i)^{v_{12}}
\left[ \left[K_{\tilde{n}'\tilde{L}'} K'_{\tilde{n}\tilde{L}}\right]_{S}
\SSS_{v_{12} S}\right]_{0} \hat{\delta}_{12}(\bm{r}'_1\bm{r}'_2;\bm{r}_1\bm{r}_2)
= i^{v_{12}}(-1)^{v_{12}+S}
\left[ \left[ K'_{\tilde{n}\tilde{L}}K_{\tilde{n}'\tilde{L}'}\right]_{S}
\SSS_{v_{12} S}\right]_{0} \hat{\delta}_{12}(\bm{r}'_1\bm{r}'_2;\bm{r}_1\bm{r}_2),
\label{eq:10appA}
\end{equation}
\end{widetext}
where the last equality results from flipping the order of coupling of the
operators $K_{\tilde{n}'\tilde{L}'}$ and  $K'_{\tilde{n}\tilde{L}}$,
which brings out the phase factor of
$(-1)^{S-\tilde{L}'-\tilde{L}}=(-1)^{S}$.
Therefore, the time-even tensor $i^{v_{12}}\hat{V}_0$ is not self-adjoint,
but we can hermitize it by using the expression given in Eq.~(\ref{eq:1}).

\section{\label{appB}Relations defining the gauge-invariant pseudopotentials}
As discussed in Section \ref{subsec:Gauge}, when the gauge
invariance is imposed on the pseudopotential, one obtains a specific set of
constraints on the parameters and terms of the pseudopotential,
which result from the condition of Eq.~(\ref{eq:5}).

At fourth order, the gauge symmetry forces seven parameters of terms listed
in the Table \ref{fourtable} to be specific linear combinations of the four
independent ones, namely,

\begin{eqnarray}
C_{00,00}^{40}&=&\frac{3}{2\sqrt{5}}C_{22,00}^{22} ,\label{subeq:1ofappB2} \\
C_{00,20}^{40}&=&\frac{3}{2\sqrt{5}}C_{22,20}^{22},\label{subeq:2ofappB2} \\
C_{00,22}^{42}&=&\frac{3}{\sqrt{7}}C_{22,22}^{22} , \label{subeq:3ofappB2} \\
C_{11,22}^{31}&=&\sqrt{\frac{21}{5}}C_{11,22}^{33} ,\label{subeq:4ofappB2} \\
C_{20,00}^{20}&=&\frac{\sqrt{5}}{2}C_{22,00}^{22} ,\label{subeq:5ofappB2}  \\
C_{20,20}^{20}&=&\frac{\sqrt{5}}{2}C_{22,20}^{22} , \label{subeq:6ofappB2} \\
C_{20,22}^{22}&=&\sqrt{7}C_{22,22}^{22} . \label{subeq:7ofappB2}
\end{eqnarray}

At sixth order, imposing the gauge symmetry forces 16 terms of the
pseudopotential listed in Table \ref{sixtable} to be specific
linear combinations of 6 independent ones, namely,

\begin{eqnarray}
C_{00,00}^{60}&=&\frac{1}{4\sqrt{5}}C_{22,00}^{42},  \label{subeq:1ofappB4}  \\
C_{00,20}^{60}&=&\frac{1}{4\sqrt{5}}C_{22,20}^{42}, \label{subeq:2ofappB4} \\
C_{00,22}^{62}&=&\frac{\sqrt{5}}{4}C_{22,22}^{44},  \label{subeq:3ofappB4}  \\
C_{11,00}^{51}&=&\frac{9}{2} \sqrt{\frac{3}{7}}C_{33,00}^{33}, \label{subeq:4ofappB4}  \\
C_{11,20}^{51}&=&\frac{9}{2} \sqrt{\frac{3}{7}}C_{33,20}^{33},\label{subeq:5ofappB4}
\end{eqnarray}
\begin{eqnarray}
C_{11,22}^{51}&=&\frac{9}{2} \sqrt{\frac{3}{35}}C_{11,22}^{53}, \label{subeq:6ofappB4}  \\
C_{20,00}^{40}&=&\frac{7}{4 \sqrt{5}}C_{22,00}^{42} ,  \label{subeq:7ofappB4} \\
C_{20,20}^{40}&=&\frac{7}{4 \sqrt{5}} C_{22,20}^{42},  \label{subeq:8ofappB4} \\
C_{20,22}^{42}&=&\frac{3\sqrt{5}}{2}C_{22,22}^{44},\label{subeq:9ofappB4}     \\
C_{22,22}^{40}&=&\frac{21}{4\sqrt{5}}C_{22,22}^{44},  \label{subeq:10ofappB4} \\
C_{22,22}^{42}&=&3 \sqrt{\frac{5}{7}}C_{22,22}^{44} ,  \label{subeq:11ofappB4} \\
C_{31,00}^{31}&=&\frac{9}{10}\sqrt{21}C_{33,00}^{33},\label{subeq:12ofappB4}    \\
C_{31,20}^{31}&=&\frac{9}{10}\sqrt{21}C_{33,20}^{33}, \label{subeq:13ofappB4}
\end{eqnarray}
\begin{eqnarray}
C_{31,22}^{31}&=&\frac{9}{10}\sqrt{\frac{21}{5}}C_{11,22}^{53},  \label{subeq:14ofappB4}  \\
C_{31,22}^{33}&=&\frac{9}{5}C_{11,22}^{53}, \label{subeq:15ofappB4}  \\
C_{33,22}^{33}&=&\sqrt{\frac{2}{15}}C_{11,22}^{53}.  \label{subeq:16ofappB4}
\end{eqnarray}

\section{\label{appD}Relations between the central-like and tensor-like pseudopotentials}

In the following we present the recoupling formulae which connect the
two alternative forms of the pseudopotential of the Eqs.~(\ref{eq:0})
and ~(\ref{eq:0t}). We have,

\begin{widetext}
\begin{eqnarray}
 \nonumber \hat{V}_{\tilde{n} \tilde{L},v_{12} S}^{\tilde{n}' \tilde{L}'}&=&
\frac{1}{2} i^{v_{12}} \left(1-\thalf\delta_{v_1,v_2}\right) \sqrt{2S+1}
 \\ \nonumber
 &\times&\left( \sum_{J=|\tilde{L}'-v_1|}^{\tilde{L}'+v_1} (-1)^{J+S+v_1+\tilde{L}}
          \sqrt{2J+1} \begin{Bmatrix} \tilde{L}' & v_1 & J \\ v_2 & \tilde{L} & S \end{Bmatrix} \left[ \left[K'_{\tilde{n}'\tilde{L}'} \sigma^{(1)}_{v_1} \right]_{J}
       \left[K_{\tilde{n}\tilde{L}} \sigma^{(2)}_{v_2} \right]_{J}\right]_{0}  \right. \\ \nonumber
&&+     \sum_{J=|\tilde{L}'-v_1|}^{\tilde{L}'+v_1} (-1)^{J+v_2+\tilde{L}} \sqrt{2J+1}
          \begin{Bmatrix} \tilde{L}' & v_1 & J \\ v_2 & \tilde{L} & S \end{Bmatrix} \left[ \left[K'_{\tilde{n}'\tilde{L}'} \sigma^{(2)}_{v_1} \right]_{J}
       \left[K_{\tilde{n}\tilde{L}} \sigma^{(1)}_{v_2} \right]_{J}\right]_{0} \\ \nonumber
&&+      \sum_{J=|\tilde{L}-v_1|}^{\tilde{L}+v_1}(-1)^{J+v_2+\tilde{L}'}\sqrt{2J+1}
         \begin{Bmatrix} \tilde{L} & v_1 & J \\ v_2 & \tilde{L}' & S \end{Bmatrix} \left[ \left[K'_{\tilde{n}\tilde{L}} \sigma^{(1)}_{v_1} \right]_{J}
       \left[K_{\tilde{n}'\tilde{L}'} \sigma^{(2)}_{v_2} \right]_{J}\right]_{0}  \\ \nonumber
&&+       \left.  \sum_{J=|\tilde{L}-v_1|}^{\tilde{L}+v_1}(-1)^{J+S+v_1+\tilde{L}'}\sqrt{2J+1}
    \begin{Bmatrix} \tilde{L} & v_1 & J \\ v_2 & \tilde{L}' & S \end{Bmatrix} \left[ \left[K'_{\tilde{n}\tilde{L}} \sigma^{(2)}_{v_1} \right]_{J}
       \left[K_{\tilde{n}'\tilde{L}'} \sigma^{(1)}_{v_2} \right]_{J}\right]_{0}   \right)                 \\
&&\times \left(1-\hat{P}^{M}\hat{P}^{\sigma}\hat{P}^{\tau}\right)
\hat{\delta}_{12}(\bm{r}'_1\bm{r}'_2;\bm{r}_1\bm{r}_2)
\label{eq:rec1}.
\end{eqnarray}

Analogously, the recoupling formula which allows to express the
tensor-like pseudopotential through the central-like one reads,
\begin{eqnarray}
 \nonumber\hat{\tilde{V}}_{\tilde{n} \tilde{L}, v_{12}J}^{\tilde{n}' \tilde{L}'} &=&
\frac{1}{2} i^{v_{12}} \left(1-\thalf\delta_{v_1,v_2}\right) \sqrt{2J+1}
\sum_{S=|\tilde{L}'-\tilde{L}|}^{\tilde{L}'+\tilde{L}} \sqrt{2S+1}
 \\ \nonumber
&\times& \left( (-1)^{S+J+v_1+\tilde{L}}  \begin{Bmatrix} \tilde{L}' & \tilde{L} & S \\ v_2 & v_1 & J \end{Bmatrix} \right.
   \left[ \left[K'_{\tilde{n}'\tilde{L}'} K_{\tilde{n}\tilde{L}} \right]_{S}
       \left[ \sigma^{(1)}_{v_1} \sigma^{(2)}_{v_2} \right]_{S}\right]_{0}
 \\ \nonumber
&& +(-1)^{J+v_2+\tilde{L}}  \begin{Bmatrix} \tilde{L}' & \tilde{L}  & S \\ v_2 & v_1 & J \end{Bmatrix}
   \left[ \left[K'_{\tilde{n}'\tilde{L}'} K_{\tilde{n}\tilde{L}} \right]_{S}
       \left[ \sigma^{(1)}_{v_2} \sigma^{(2)}_{v_1} \right]_{S}\right]_{0}
 \\ \nonumber
&& + (-1)^{S+J+v_1+\tilde{L}'} \begin{Bmatrix} \tilde{L} & \tilde{L}' & S \\ v_2 & v_1 & J \end{Bmatrix}
   \left[ \left[K'_{\tilde{n}\tilde{L}} K_{\tilde{n}'\tilde{L}'} \right]_{S}
       \left[ \sigma^{(1)}_{v_1} \sigma^{(2)}_{v_2} \right]_{S}\right]_{0}
  \\ \nonumber
&& \left. + (-1)^{J+v_2+\tilde{L}'} \begin{Bmatrix} \tilde{L} & \tilde{L}'& S \\ v_2 & v_1  & J \end{Bmatrix}
   \left[ \left[K'_{\tilde{n}\tilde{L}} K_{\tilde{n}'\tilde{L}'} \right]_{S}
       \left[ \sigma^{(1)}_{v_2} \sigma^{(2)}_{v_1} \right]_{S}\right]_{0}  \right)                 \\
&&\times \left(1-\hat{P}^{M}\hat{P}^{\sigma}\hat{P}^{\tau}\right)
\hat{\delta}_{12}(\bm{r}'_1\bm{r}'_2;\bm{r}_1\bm{r}_2)
\label{eq:rec2}.
\end{eqnarray}
\end{widetext}

According to the recoupling of the Eq.~(\ref{eq:rec1}), we give the list of the relations between the parameters of the two forms of the pseudopotential.
For the second order terms we have,

\begin{eqnarray}
C_{00,00}^{20}&=&\tilde{C}_{00,00}^{20} ,  \label{subeq:1of10} \\
C_{00,20}^{20}&=&\tilde{C}_{00,21}^{20},  \label{subeq:2of10} \\
C_{00,22}^{22}&=&\tilde{C}_{00,21}^{22}, \label{subeq:3of10} \\
C_{11,00}^{11}&=&\tilde{C}_{11,01}^{11}, \label{subeq:4of10} \\
C_{11,20}^{11}&=&\frac{1}{3}\tilde{C}_{11,20}^{11}+\frac{1}{\sqrt{3}}\tilde{C}_{11,21}^{11}+\frac{\sqrt{5}}{3}\tilde{C}_{11,22}^{11} , \label{subeq:5of10} \\
C_{11,11}^{11}&=&-\tilde{C}_{11,11}^{11} , \label{subeq:6of10} \\
C_{11,22}^{11}&=&\frac{\sqrt{5}}{3}\tilde{C}_{11,20}^{11}-\frac{\sqrt{5}}{2\sqrt{3}}\tilde{C}_{11,21}^{11}+\frac{1}{6}\tilde{C}_{11,22}^{11}  \label{subeq:7of10};
\end{eqnarray}

at the fourth order,

\begin{equation}
 C_{00,00}^{40}=\tilde{C}_{00,00}^{40} , \label{subeq:1of11}  \\
\end{equation}
\begin{equation}
C_{00,20}^{40}= \tilde{C}_{00,21}^{40}                         ,   \label{subeq:2of11}  \\
\end{equation}
\begin{equation}
 C_{00,22}^{42}= \tilde{C}_{00,21}^{42}                         ,   \label{subeq:3of11}  \\
\end{equation}
\begin{equation}
 C_{11,00}^{31}= \tilde{C}_{11,01}^{31}                         ,    \label{subeq:4of11} \\
\end{equation}
\begin{equation}
 C_{11,20}^{31}= \frac{1}{3}\tilde{C}_{11,20}^{31} + \frac{1}{\sqrt{3}}\tilde{C}_{11,21}^{31} + \frac{\sqrt{5}}{3}\tilde{C}_{11,22}^{31} ,    \label{subeq:5of11}  \\
\end{equation}
\begin{equation}
 C_{11,11}^{31}= -\tilde{C}_{11,11}^{31}   ,   \label{subeq:6of11}  \\
\end{equation}
\begin{equation}
 C_{11,22}^{31}= \frac{\sqrt{5}}{3}\tilde{C}_{11,20}^{31} -\frac{\sqrt{5}}{2\sqrt{3}} \tilde{C}_{11,21}^{31} + \frac{1}{6}\tilde{C}_{11,22}^{31}     ,   \label{subeq:7of11} \\
\end{equation}
\begin{equation}
 C_{11,22}^{33}= \tilde{C}_{11,22}^{33}                         ,  \label{subeq:8of11}  \\
\end{equation}
\begin{equation}
 C_{20,00}^{20}= \tilde{C}_{20,00}^{20}                         ,  \label{subeq:9of11}  \\
\end{equation}
\begin{equation}
 C_{20,20}^{20}= \tilde{C}_{20,21}^{20}                         ,   \label{subeq:10of11} \\
\end{equation}
\begin{equation}
 C_{20,22}^{22}= \tilde{C}_{20,21}^{22}                         ,  \label{subeq:11of11}  \\
\end{equation}
\begin{equation}
 C_{22,00}^{22}= \tilde{C}_{22,02}^{22}                         ,  \label{subeq:12of11}  \\
\end{equation}
\begin{equation}
 C_{22,20}^{22}= \frac{1}{\sqrt{5}}\tilde{C}_{22,21}^{22}+\frac{1}{\sqrt{3}}\tilde{C}_{22,22}^{22} +  \sqrt{\frac{7}{15}}\tilde{C}_{22,23}^{22}                       ,  \label{subeq:13of11}  \\
\end{equation}
\begin{equation}
 C_{22,11}^{22}= -\tilde{C}_{22,12}^{22}                         ,   \label{subeq:14of11} \\
\end{equation}
\begin{equation}
 C_{22,22}^{22}= \frac{\sqrt{35}}{10}\tilde{C}_{22,21}^{22}-\frac{\sqrt{7}}{2\sqrt{3}}\tilde{C}_{22,22}^{22} +  \frac{1}{\sqrt{15}}\tilde{C}_{22,23}^{22};   \label{subeq:15of11}
\end{equation}

at the sixth order,
\begin{equation}
 C_{00,00}^{60}=\tilde{C}_{00,00}^{60} , \label{subeq: 1of12}  \\
\end{equation}
\begin{equation}
 C_{00,20}^{60}=\tilde{C}_{00,21}^{60} , \label{subeq: 2of12} \\
\end{equation}
\begin{equation}
 C_{00,22}^{62}=\tilde{C}_{00,21}^{62} , \label{subeq: 3of12}  \\
\end{equation}
\begin{equation}
 C_{11,00}^{51}=\tilde{C}_{11,01}^{51} , \label{subeq: 4of12} \\
\end{equation}
\begin{equation}
 C_{11,20}^{51}= \frac{1}{3}\tilde{C}_{11,20}^{51} + \frac{1}{\sqrt{3}}\tilde{C}_{11,21}^{51} +\frac{\sqrt{5}}{3}\tilde{C}_{11,22}^{51}, \label{subeq: 5of12} \\
\end{equation}
\begin{equation}
 C_{11,11}^{51}= -\tilde{C}_{11,11}^{51} , \label{subeq: 6of12} \\
\end{equation}
\begin{equation}
 C_{11,22}^{51}= \frac{\sqrt{5}}{3}\tilde{C}_{11,20}^{51} -\frac{\sqrt{5}}{2\sqrt{3}} \tilde{C}_{11,21}^{51} + \frac{1}{6}\tilde{C}_{11,22}^{51} ,  \label{subeq: 7of12} \\
\end{equation}
\begin{equation}
 C_{11,22}^{53}=\tilde{C}_{11,22}^{53} ,  \label{subeq: 8of12} \\
\end{equation}
\begin{equation}
 C_{20,00}^{40}=\tilde{C}_{20,00}^{40} ,  \label{subeq: 9of12} \\
\end{equation}
\begin{equation}
 C_{20,20}^{40}=\tilde{C}_{20,21}^{40} , \label{subeq: 10of12}  \\
\end{equation}
\begin{equation}
 C_{20,22}^{42}=\tilde{C}_{20,21}^{42} , \label{subeq: 11of12}  \\
\end{equation}
\begin{equation}
 C_{22,22}^{40}=\tilde{C}_{22,21}^{40} ,  \label{subeq: 12of12} \\
\end{equation}
\begin{equation}
 C_{22,00}^{42}=\tilde{C}_{22,02}^{42} ,  \label{subeq: 13of12} \\
\end{equation}
\begin{equation}
 C_{22,20}^{42}= \frac{1}{\sqrt{5}}\tilde{C}_{22,21}^{42}+\frac{1}{\sqrt{3}}\tilde{C}_{22,22}^{42} +  \sqrt{\frac{7}{15}}\tilde{C}_{22,23}^{42} ,  \label{subeq: 14of12} \\
\end{equation}
\begin{equation}
 C_{22,11}^{42}=-\tilde{C}_{22,12}^{42} ,   \label{subeq: 14bisof12} \\
\end{equation}
\begin{equation}
 C_{22,22}^{42}= \frac{\sqrt{35}}{10}\tilde{C}_{22,21}^{42}-\frac{\sqrt{7}}{2\sqrt{3}}\tilde{C}_{22,22}^{42} +  \frac{1}{\sqrt{15}}\tilde{C}_{22,23}^{42},  \label{subeq: 15of12} \\
\end{equation}
\begin{equation}
 C_{22,22}^{44}=\tilde{C}_{22,23}^{44} ,  \label{subeq: 16of12}  \\
\end{equation}
\begin{equation}
 C_{31,00}^{31}=\tilde{C}_{31,01}^{31} ,  \label{subeq: 17of12}   \\
\end{equation}
\begin{equation}
 C_{31,20}^{31}= \frac{1}{3}\tilde{C}_{31,20}^{31} + \frac{1}{\sqrt{3}}\tilde{C}_{31,21}^{31} +\frac{\sqrt{5}}{3}\tilde{C}_{31,22}^{31} , \label{subeq: 18of12} \\
\end{equation}
\begin{equation}
 C_{31,11}^{31}=-\tilde{C}_{31,11}^{31} ,  \label{subeq: 19of12}  \\
\end{equation}
\begin{equation}
 C_{31,22}^{31}=\frac{\sqrt{5}}{3}\tilde{C}_{31,20}^{31} -\frac{\sqrt{5}}{2\sqrt{3}} \tilde{C}_{31,21}^{31} + \frac{1}{6}\tilde{C}_{31,22}^{31} ,  \label{subeq: 20of12}  \\
\end{equation}
\begin{equation}
 C_{31,22}^{33}=\tilde{C}_{31,22}^{33} ,   \label{subeq: 21of12}  \\
\end{equation}
\begin{equation}
 C_{33,00}^{33}=\tilde{C}_{33,03}^{33} ,  \label{subeq: 22of12}  \\
\end{equation}
\begin{equation}
 C_{33,20}^{33}=\sqrt{\frac{5}{21}}\tilde{C}_{33,22}^{33}+\frac{1}{\sqrt{3}}\tilde{C}_{33,23}^{33}+\sqrt{\frac{3}{7}}\tilde{C}_{33,24}^{33},  \label{subeq: 23of12}  \\
\end{equation}
\begin{equation}
 C_{33,11}^{33}=-\tilde{C}_{33,13}^{33} ,  \label{subeq: 24of12}  \\
\end{equation}
\begin{equation}
 C_{33,22}^{33}=\sqrt{\frac{2}{7}}\tilde{C}_{33,22}^{33}-\frac{\sqrt{5}}{2\sqrt{2}}\tilde{C}_{33,23}^{33}+\frac{\sqrt{5}}{2\sqrt{14}}\tilde{C}_{33,24}^{33}.   \label{subeq: 25of12}
\end{equation}


\end{document}